\documentclass[twocolumn,showpacs,preprintnumbers,amsmath,amssymb,superscriptaddress]{revtex4-1}

\usepackage{graphicx}
\usepackage{dcolumn}
\usepackage{bm}
\usepackage{color}
\usepackage{pifont}
\usepackage{natbib}
\usepackage{hyperref}
\hypersetup{
colorlinks = true,
urlcolor   = blue,
linkcolor  = blue,
citecolor  = blue
}

\begin{document}

\title{Refined magnetic structure of VI$_3$}

\author{Ola~Kenji~Forslund}
 \email{olake@chalmers.se}
\affiliation{Department of Physics, Chalmers University of Technology, SE-412 96 G\"oteborg, Sweden}

\author{Yuqing~Ge}
\affiliation{Department of Physics, Chalmers University of Technology, SE-412 96 G\"oteborg, Sweden}

\author{Hiroto~Ohta}
\affiliation{Faculty of Science and Engineering, Doshisha University, Kyotanabe, Kyoto 610-0321, Japan}

\author{Chennan Wang}
\affiliation{Paul Scherrer Institute, Laboratory for Muon Spin Spectroscopy, CH-5232 PSI Villigen, Switzerland}

\author{Mahmoud~Abdel-Hafiez}
\affiliation{Department of Physics and Astronomy, Uppsala University, Ångströmlaboratoriet, SE-75120 Uppsala, Sweden}

\author{Jun~Sugiyama}
\affiliation{Neutron Science and Technology Center, 
Comprehensive Research Organization for Science and Society (CROSS), Tokai, Ibaraki 319-1106, Japan}

\author{Martin~M\aa nsson}
\affiliation{Department of Applied Physics, KTH Royal Institute of Technology, SE-106 91 Stockholm, Sweden}

\author{Yasmine Sassa}
\email{yasmine.sassa@chalmers.se}
\affiliation{Department of Physics, Chalmers University of Technology, SE-412 96 G\"oteborg, Sweden}

\date{\today}

\begin{abstract}
The van der Waals ferromagnet (FM), VI$_3$, was studied by muon spin relaxation ($\mu^+$SR) and first principle calculations based on density functional theory (DFT). Temperature dependent zero field muon spin relaxation ($\mu^+$SR) measurements confirm the onset of long range FM order and the time spectra exhibits clear muon spin precession frequencies for $T<T_{\rm C}=50.03(1)$~K. The calculated internal magnetic fields at the predicted muon sites, based on the established magnetic structure from neutron diffraction, is inconsistent with the measured one. This inconsistency is because of strong incoherent neutron scattering and absorption originating from the elements V and I. Instead, a new and a more accurate magnetic structure is derived based on a combined study using $\mu^+$SR and DFT. These results suggest strong contritions from orbital angular momentum, providing experimental evidence for the existence of unquenched orbital angular momentum of V$^{3+}$ in VI$_3$. Finally, an unusual form of a short range ordering is present above $T_{\rm C}$. Its temperature dependence is unlike previously reported cases in other layered compounds and its microscopic origin is discussed.

\end{abstract}



\maketitle

\section{\label{sec:Intro}Introduction}
The linear spin-chain Ising model is the foundation behind the now established field, low dimensional magnetism \cite{Ising1924}. The premise behind low dimensional magnetism is having exchange interactions restricted in one or two spatial directions, effectively realising a low dimensional magnet even in three dimensional materials. CrX$_3$ ($X =$~I, Cl) is a series of compounds exhibiting a low dimensional character. These materials are stabilised under a rhombohedral symmetry, consisting of 2D Cr layers arranged in a honeycomb web fashion, surrounded by octahedrally coordinated $X$ ions.

Many studies on Cr$X_3$ were performed back in the 60s, but the materials have regained attention due to current topical interest in 2D layered materials. Notably a recent study has shown that monolayers may be obtained via exfoliation \cite{Huang2017}. The interest is driven by the possibility to study low dimensional magnetism, and prospects of electrical control of magnetism for future functional devices \cite{Matsukura2015, Jiang2018}. 
The recent spike in the interests in these van der Waals magnets have resulted into the emergence of a new compound belonging to the same family: VI$_3$ \cite{Kong2019, Tian2019, Son2019, Doleifmmode2019, Marchandier2021}. While CrI$_3$ is in Cr$^{3+}$ configuration with half-filled t$_{2g}$ levels with $S = 3/2$, VI$_3$ is in V$^{3+}$ configuration with $S = 1$. Similar to the CrI$_3$, initial magnetization measurements suggested VI$_3$ to be a ferromagnetic (FM) compound with $T_{\rm C} = 49$~K. The measured magnetic susceptibility shows a Curie-Weiss temperature dependence with an effective moment of $2.6~\mu_{\rm B}$ per V and is consistent with spin-1 configuration of V$^{3+}$. Around $T_S = 80$~K, the system undergoes a structural transition from trigonal symmetry at $T > T_S$ to a monoclinic solution at $T < T_S$.

A very recent report based on ac susceptibility measurements hinted at the existence of multiple transitions in VI$_3$ \cite{Valenta2021}. In detail, $T_1 = 54.5$~K, $T_2 =53$~K, $T_{\rm C} = 49.5$~K and $T_{\rm FM} =26$~K where reported. $T_{\rm C}$ corresponds to the previously reported long range FM order while $T_{\rm FM}$ is characterized by a drop in the AC susceptibility and was attributed to FM structure reorientation \cite{Gati2019} due to a structural transition from monoclinic symmetry to a triclinic symmetry at low temperatures \cite{Doleifmmode2019}. $T_1$ and $T_2$ on the other hand are somewhat ambiguous. Although, the detailed nature remains unsolved but one suggestion includes surface degradation.

\begin{figure*}[ht]
  \begin{center}
     \includegraphics[keepaspectratio=true,width=\textwidth]{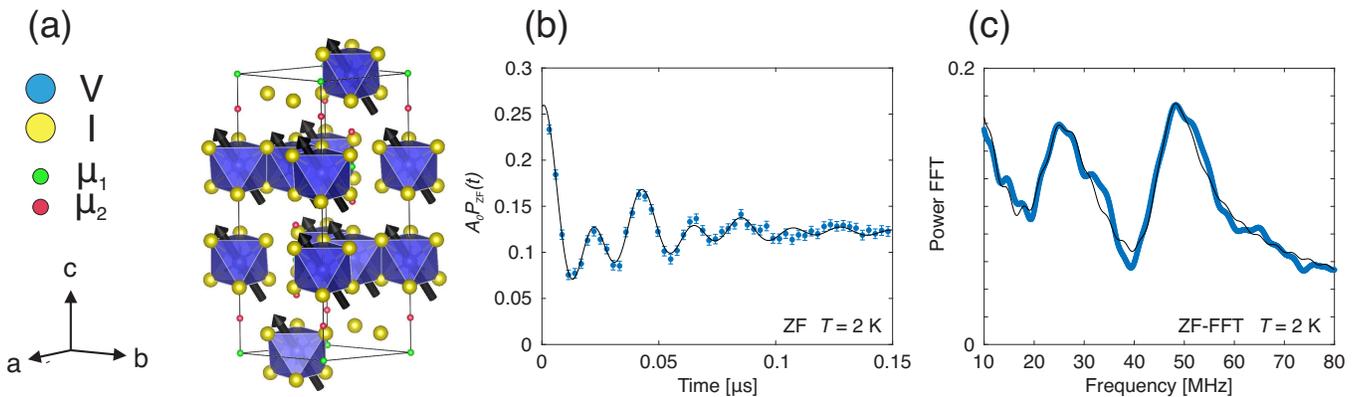}
 \end{center}
    \caption{(a) The crystal structure of VI$_3$ under rhombohedral symmetry. The predicted muon sites are included with at the positions (0,0,0) (green) and (0,0,0.125) (red). (b) The Zero field $\mu^+$SR time spectra collected at $T=2$~K. (c) Power Fourier transform of (b). The solid lines in (b) and (c) are best fits obtained using Eq.~(\ref{eq:ZF}).}
    \label{fig:structure}
\end{figure*}

Formation of massless Dirac magnons were observed in the van der Waals magnets CrI$_3$ \cite{Chen2018, Chen2020}, CrBr$_3$ \cite{Cai2021} and CrCl$_3$ \cite{Chen2022, Do2022}. Topological gaps were reported for CrI$_3$ and CrBr$_3$, consistent with having the FM order perpendicular to the honeycomb crystal structure. The Dirac points in CrCl$_3$ were on the other hand not gapped because of the in plane AF structure. Although, a recent $\mu^+$SR measurement \cite{Forslund2021} suggested the possibility of canted AF structure in CrCl$_3$, which together with the increase in internal field dynamics at lower temperatures may suggest that the Dirac points are actually gapped and was just not fully resolvable \cite{Do2022}. Even though the honeycomb crystal structure is distorted for VI$_3$ in the FM phase, the magnetic structure is a quantify that defines the physics in these systems. Neutron diffraction (ND) is the most straight forward and suitable technique to determine the magnetic structure. However, the strong incoherent scattering from V and absorption from I make such determination difficult for VI$_3$ \cite{Hao2021, Marchandier2021}. Instead, we propose to clarify the magnetic structure in this material via zero field (ZF) $\mu^+$SR combined with DFT calculations. This work highlights $\mu^+$SR, combined with DFT calculations, as a viable choice in determining the magnetic structures in magnetic materials.

\section{\label{sec:exp}Experimental Setup}
The $\mu^+$SR measurements were performed at the surface muon beamline GPS at PSI (Switzerland). Approximately, 700~mg of VI$_3$ was prepared inside a thin ($\simeq50~\mu$m) aluminium coated mylar envelope. This envelope was sealed inside a He glove box and closed with a kapton tape. This envelope was mounted on a low background Cu fork sample stick and inserted into a $^4$He flow cryostat to reach $T_{\rm base}\simeq 2$~K. The software package \texttt{musrfit} was used in order to analyze the data \cite{musrfit}. 

The muon sites were predicted with DFT calculation using a pseudopotential-based plane-wave method as implemented in $Quantum~Espresso$ \cite{QE-2009, QE-2017}. The psudopotentials were taken from Refs.~\cite{Lejaeghere2016, Prandini2018} and the muon sites were assumed to be situated in an electrostatic minimum, determined via a self consistent run. Any local distortion due to the implanted muon were not considered.

\section{\label{sec:results}Results}
Zero field (ZF) time spectrum and the corresponding Fourier transform, collected at $T = 2$~K is shown in Fig.~\ref{fig:structure}(b, c). Two clear oscillations are clearly visible in the figures, consistent with the formation of ferromagnetic (FM) order with two muon sites. As the temperature increases, the oscillations are turned into an exponential and a Gaussian Kubo-Toyabe (G-KT) like polarisation. In order to account for the processes observed for the considered temperature range, the ZF time spectra were fitted using a combination of cosine functions, exponentials and G-KT:

\begin{eqnarray}
 A_0 \, P_{\rm ZF}(t) &=&
\sum_i^2 A_{\rm FMi} \cos(f_{\rm FMi}\pi t+\phi_i)e^{-\lambda_{\rm FMi}t} + A_{\rm tail}e^{-(\lambda_{\rm tail} t)^\beta} \cr & + & A_{\rm F}e^{-\lambda_{\rm F} t} + A_{\rm S}e^{-\lambda_{\rm S} t} \cr & + &A_{\rm KT}G^{SGKT}(t,\Delta_{\rm KT})e^{-\lambda_{\rm KT} t},
\label{eq:ZF}
\end{eqnarray}

where $A_0$ is the total asymmetry of the experimental setup and $P_{\rm ZF}(t)$ is the ZF depolarisation function describing the system under study. $A_{\rm FMi}$, $f_{i}$, $\phi_i$ and $\lambda_{\rm FMi}$ are the asymmetry, precession frequency, phase and relaxation rate of the oscillating signals while $A_{\rm tail}$ and $\lambda_{\rm tail}$ are the asymmetry and relaxation rate of the tail components of the oscillating signal. $\beta$ is the stretched component of the tail and entails a distributions of spin-spin correlation times. In a powder sample, 1/3 of the internal field components are parallel to the initial muon spin polartisation (the tail component: $A_{\rm tail}$) and 2/3 of the internal field components are perpendicular to the initial muon spin polarisation (the oscillating components $\sum A_{\rm FMi}$). 

\begin{figure}[ht]
  \begin{center}
     \includegraphics[keepaspectratio=true,width=79 mm]{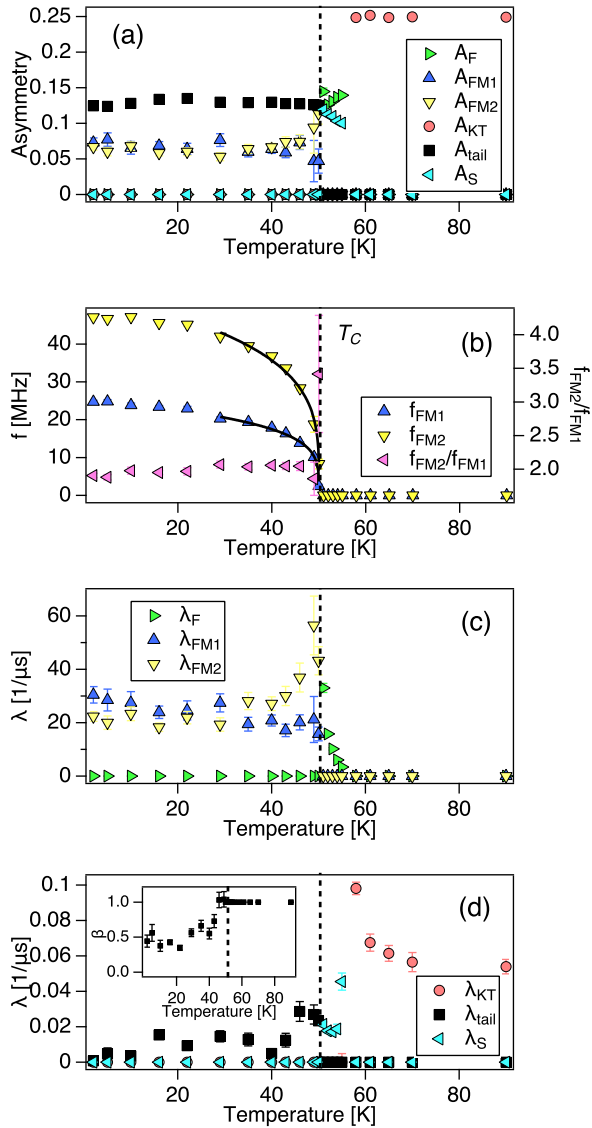}
 \end{center}
    \caption{Temperature dependent zero field (ZF) fit parameters obtained suing Eq.~(\ref{eq:ZF}): (a) asymmetry ($A_{\rm FM1}$, $A_{\rm FM2}$, $A_{\rm tail}$, $A_{\rm F}$, $A_{\rm S}$ and $A_{\rm KT}$), (b) presession frequencies ($f_{\rm FM1}$ and $f_{\rm FM2}$), (c) fast relaxation rates ($\lambda_{\rm FM1}$, $\lambda_{\rm FM2}$ and $\lambda_{\rm F}$) and (d) slow relaxation rates ($\lambda_{\rm tail}$, $\lambda_{\rm S}$ and $\lambda_{\rm KT}$). The inset in (d) show the stretched exponent of the tail. The solid lines in (b) is a fit using the expression $f(T) = f_0(1-T/T_{\rm C})^\alpha$; the fit parameters are listed in the main text.}
    \label{fig:ZF_para}
\end{figure}

$A_{\rm F}$, $\lambda_{\rm F}$ $A_{\rm S}$ and $\lambda_{\rm S}$ are the asymmetries and relaxation rates of a fast (F) and slow (S) components, respectively. $A_{\rm KT}$, $\lambda_{\rm KT}$ and $\Delta_{\rm KT}$ are the asymmetry, relaxation rate and field distribution width of the KT contribution. Since weak longitudinal field is able to decouple decouple the spectra at higher temperatures, the random fields which is the origin to the KT component is confirmed to be originating from $I^{50}_{\rm V} = 6$, $I^{51}_{\rm V} = 7/2$ and $I^{127}_{\rm I}=5/2$ nuclear spins. The origin of the two exponentials ($A_{\rm F}$ and $A_{\rm S}$) is discussed below. It is noted that the components of the first row of Eq.~(\ref{eq:ZF}), $A_{\rm tail}$ and $A_{\rm FMi}$, are only none zero for $T\leq T_{\rm C}$, the second row ($A_{\rm F}$ and $A_{\rm S}$) is none zero for $T_{\rm C} < T \leq 58$~K and $A_{\rm KT}$ is none zero for $T>58$~K.

Since the compound is known to be a ferromagnet, $\phi_i = \phi$ was set in the fitting procedure. The temperature dependencies of the ZF fit parameters are shown in Fig.~\ref{fig:ZF_para}. The precession frequencies ($f_{\rm FMi}$) show an order parameter like behaviour. A fit with the mean field expression, $f(T) = f_0(1-T/T_{\rm C})^\alpha$, close to the critical temperature yields $f_0 = 25.4(6)$, $T_{\rm C} = 50.012(7)$ and $\alpha=0.232(12)$ for the $f_{\rm FM1}$ and $f_0 = 53.8(1.7)$, $T_{\rm C} = 50.04(3)$ and $\alpha=0.255(18)$ for $f_{\rm FM2}$. Since $\alpha$ and $T_{\rm C}$ should be muon site independent, the average value of the two sites defines $T_{\rm C}= 50.03(1)$ and $\alpha=0.244(11)$. The value of the critical exponent is consistent with an XY magnet \cite{Taroni2008}. In fact, a recent single crystal inelastic neutron scattering (INS) measurement confirms none or only small dispersive behaviour along the $c$-axis, indicative of a 2D behaviour \cite{Lane2021}. 

Previous XRD, magnetic susceptibility and heat capacity studies \cite{Son2019, Doleifmmode2019, Marchandier2021} have suggested a structural transition to occur around 32~K. This distortion resulted into a spin reorientation of the FM order \cite{Hao2021}. However, transitions or anomalies are not observed around these temperatures in the presented data (Fig.~\ref{fig:ZF_para}). Since a spin reorientation is expected to alter the local magnetic field at the muon sites, the precession frequencies (Fig.~\ref{fig:ZF_para}(b)) are expected to change. Moreover, the obtained value of $f_0$ (the precession frequency at base temperature) is inconsistent with the reported magnetic structure by neutron diffraction \cite{Hao2021, Marchandier2021}. These points are further discussed in Sec.~\ref{sec:discussion}. 

The obtained value $A_{\rm tail}\simeq A_0/2$ is slightly higher than expected ($A_{\rm tail}\simeq A_0/3$). Since the samples are flake like, it is likely that preferential orientation is present such that the out of plane ordered moments orient parallel to the initial muon spin polarisation. This assessment consistent with the solved magnetic structure (Fig.~\ref{fig:structure}(a)), in which the magnetic moments have a significant out of plane component. Above 58~K, exponentially relaxing $A_{\rm KT}$ covers the full asymmetry and is indicative of formation of a paramagnetic state. The intermediate temperatures on the other hand, $T_{\rm C}\leq T<58$~K, are described by two exponentials, a fast ($A_{\rm F}$) and a slow ($A_{\rm S}$) one. These exponentials seem to be originating from the oscillation, $A_{\rm FM2}$, that are damped into a fast relaxing component ($A_{\rm F}$) and the tail ($A_{\rm tail}$) which is transformed into a slow component ($A_{\rm S}$) at $T_{\rm C}$. These behaviours are typical for the formation of a short range order and was recently shown to also be the case for the similar compound CrCl$_3$ \cite{Forslund2021}. In fact, other similar layered compounds show similar SRO formation just above $T_C$ \cite{Baker2005, Forslund_2020_Na, Forslund_2021_La}. The difference in this case is however that as the temperature increases in the region $T_{\rm C}\leq T<58$~K, the fast component increase in asymmetry while the slow one decrease. In all other reported case, to the best of our knowledge, it is the opposite in which the fast component decreases in asymmetry while the slow component increases. Such behaviour is based on the fact that perpendicular magnetic fluctuations decreases with increasing temperature, while longitudinal one increases. The origin of this behaviour is discussed in Sec.~\ref{sec:discussion}.

The relaxation rates of the oscillating components ($\lambda_{\rm FMi}$) show a temperature independent behaviour at low temperatures. $\lambda_{\rm FMi}$ corresponds roughly to the spin-spin relaxation rate and indicates the field distribution width at the muon site. Close to $T_{\rm C}$, one of the component increases, as commonly observed close to the critical temperature. $\lambda_{\rm tail}$ represent the spin lattice relaxation rate and corresponds to the internal field dynamics. A linear increase is observed up to about 15~K and a platue up to 35~K, which is followed by another increase just below $T_{\rm C}$. These temperatures corresponds to the reported structural transition and suggest that the structural transition are associated with internal field dynamics. In the intermediate temperature range, $T_{\rm C}\leq T<58$~K, $\lambda_{\rm F}$ exhibits a sharp decrease $\lambda_{\rm S}$ increases. Finally, $\lambda_{\rm KT}$ increase close to $T_{\rm C}$ and its temperature dependence is most likely a Curie-Weiss behaviour at higher temperature.

While the structural transition at 32~K is not observed in the precession frequencies, the stretched exponent of the spin-lattice relaxation rate ($\beta$) start to increase around 30~K. Since $\lambda_{\rm tail}$ is proportional to the dynamics in the system, it is noted that the structural distortion is associated with local field fluctuations. Fluctuation driven structural distortion is commonly observed in many systems exhibiting second order transitions. 

\begin{table*}[ht]
\caption{\label{table:Fre1}
The calculated local field ($f_{loc}$) values at the predicted muon sites are tabulated together with the obtained local spin density $\rho(\bm r_{\mu})$ for the proposed magnetic structure by Ref.~\onlinecite{Hao2021}. 
$f_{\rm loc}$ is given by $f_{loc} =\frac{\gamma_{\mu}}{2\pi} | \bm B_{\rm dip'} + \bm B_{\rm L} + \bm B_{\rm hf} |$ where $\bm B_{\rm dip'}= \frac{\mu_0}{4\pi}\sum^{N}_{j}\frac{3\bm r_{\mu j}(\bm m_{e,j}\cdot\bm r_{\mu j})}{r_{\mu j}^5}-\frac{\bm m_{e,j}}{r_{\mu j}^3}$, $\bm B_{\rm L}=\frac{\mu_0}{3}\bm M_{\rm L}=\frac{\mu_0}{3V}\sum^N_j \bm m_{e,j}$ and $\bm B_{\rm hf}=\frac{2\mu_0}{3}| \psi(\bm r_\mu) | ^2 \bm m_e=\frac{2\mu_0}{3} \frac{\rho(\bm r_{\mu})}{| \bm m_e | }\bm m_e$. Here, $\bm m_e$ is the magnetic moment of the electron and $\bm r_{\mu, j}$ is the distance between the muon and the j-th atom. The experimentally determined base temperature precession frequency $f_0$ is listed as well. The calculation was performed using the python package $MUESR$ \cite{Bonfa2017}.
}

\begin{ruledtabular}
\begin{tabular}{lcccccc}
 Muon site & $\rho(\bm r_{\mu})~[\mu_{\rm B}$\AA$^{-3}]$&$\bm B_{\rm dip'}$~[T]&$\bm B_{\rm hf}$~[T]&$\bm B_{\rm L}$~[T]&$f_{\rm loc}$~[MHz] &$f_0$~[MHz]\\
 \hline
 $\mu1$~(0,0,0) & 0.000306552 &[0,0.0299,-0.0769]&[0,0.0007,0.0009]&[0,0.0236,0.0304]&9.60405&  53.8(1.7)\\

 \hline
 $\mu2$~(0,0,0.125) & -0.000117653
 &[0,-0.0096,0.0246] &[0,-0.0003,-0.0004]&[0,0.0236,0.0304] &7.63057&    25.39(55)\\
\end{tabular}
\end{ruledtabular}
\end{table*}

\begin{table*}[ht]
\caption{\label{table:Fre2}
The calculated local field ($f_{loc}$) values at the predicted muon sites are tabulated, together with the obtained local spin density $\rho(\bm r_{\mu})$, for the tiled and scaled magnetic structure (Fig.~\ref{fig:structure}(a), see main text). Beside the assumed magnetic structure, the calculations procedure is the same to table~\ref{table:Fre1}.
}

\begin{ruledtabular}
\begin{tabular}{lcccccc}
 Muon site & $\rho(\bm r_{\mu})~[\mu_{\rm B}$\AA$^{-3}]$&$\bm B_{\rm dip'}$~[T]&$\bm B_{\rm hf}$~[T]&$\bm B_{\rm L}$~[T]&$f_{\rm loc}$~[MHz] &$f_0$~[MHz]\\
 \hline
 $\mu1$~(0,0,0) & 0.000306552 &[0,0.2043,-0.2172]&[0,0.0048,0.0026]&[0,0.0048,0.0026]&53,17&  53.8(1.7)\\

 \hline
 $\mu2$~(0,0,0.125) & -0.000117653
 &[0,-0.0653,0.0694] &[0,0.0019,-0.001]&[0,0.1614,0.0858] &24.50&    25.39(55)\\
\end{tabular}
\end{ruledtabular}
\end{table*}

\section{\label{sec:discussion} Discussion}
We shall first address the question to why any traces of the structural transition around 32~K is absent in the muon data. First of all, it is noted that muons are susceptible to magnetic fields. Structural transition are in principle not directly visible with $\mu^+$SR. In this case however, the structural distortion was accompanied by a magnetic structure reorientation. Powder neutron diffraction measurements reported two different ferromagnetic structure below and above 30~K \cite{Hao2021}. However, the difference in the magnetic order between the two (apart from the small structure distortion) consists of a rotation within the ab plane by 90$^\circ$. Assuming that the underlined crystal structure is hexagonal, this kind of rotation is not expected to alter the local magnetic field since a and b axis are equivalent. If a lower crystal symmetry is assumed, this kind of rotation will affect the local field and thus the precession frequency. However, since any changes in the local field is not observed at this temperature (Fig.~\ref{fig:ZF_para}(b)), we may conclude that the reported structural distortion is small such that the system can be approximated under hexagonal structure even at low temperature. This kind of conclusion is consistent with the reported inelastic neutron scattering study \cite{Lane2021}. 

In order to deduce the magnetic structure, the crystalline muon sites were calculated under the assumption of a hexagonal crystal structure. Assuming that the electrostatic minimum corresponds to a muon site, we propose two sites: (0,0,0) and (0,0,0.125) (Fig.~\ref{fig:structure}(a)). In fact, the distance between the V$^{3+}$ ions and the predicted muon sites are too large to account for the observed precession frequency (Fig.~\ref{fig:ZF_para}(b)), assuming an ordered moment determined from ND ($1.3(1)~\mu_{\rm B}$ \cite{Hao2021} or $0.2~\mu_{\rm B}$ \cite{Marchandier2021}) . In fact, local field calculations based on the magnetic structure determined by ND and the predicted muon sites yields presession frequencies of about 7 and 9~MHz (table~\ref{table:Fre1}). These frequencies are significantly lower from the experientially observed values (Fig.~\ref{fig:ZF_para}(b)). Inconsistency between ND and $\mu^+$SR has been observed in other compounds \cite{Nozaki2018, Potashnikov2021}. The system was either itinerant and it was speculated that the two different techniques are sensitive to fluctuations at different time scales or the samples were containing elements that are highly absorbing neutrons. Since VI$_3$ is a semi-conductor, itinerant moments are likely to play less of a role. Instead, the inconsistency has two possible explanations; 1) the predicted muon sites are not correct. While this scenario is possible, it is not very likely since similar muon sites were concluded and confirmed for the isotructural compound CrCl$_3$ \cite{Forslund2021}). Moreover, the muons need to be put inside the VI$_6$ octahedra in order to achieve local field values consistent with experiment (not realistic) \cite{Sulaiman1994}. 2) The determined magnetic structure from ND is not reliable. Here, we shall consider the following why this may be the case; the magnetic signal will overlap with the crystal structure and separating each component can in certain cases be difficult for a FM compound. Moreover, V is a strong incoherent scatterer and I is a strong absorber of neutrons. These facts distorts the data and intrinsic values are thus not obtained. Therefore, it is highly likely that the magnetic structure determined from ND is not accurate.

Following upon our previous works \cite{Forslund_2020_Na, Forslund_2021_La, Forslund2020_La}, we shall estimate the magnetic structure based on the presented data. In detail, the expected local magnetic field (see table~\ref{table:Fre1} and table~\ref{table:Fre2}), established from a given magnetic structure, is calculated at the predicted muon sites ((0,0,0) and (0,0,0.125)). A clear discrepancy is observed between the calculated value ($f_{\rm loc}$) and the experimental value ($f_0$). To obtain better results, we may increase the ordered moments. However, we shall note that $\frac{f^{\mu1}_{\rm loc}}{f^{\mu2}_{\rm loc}}\simeq1.26$ while $\frac{f^{\mu1}_0}{f^{\mu2}_0}\simeq2.12$ (Fig.~\ref{fig:ZF_para}(b)). Therefore, a simple scaling of ordered moments cannot fully explain the discrepancy. Indeed, scaling ordered magnetic moments to $\mu_{\rm ord}\sim4.4~\mu_{\rm B}$ yields $f_{\rm loc}=25.18$ and 31.69~MHz for the sites $\mu_2$ and $\mu_1$, respectively. Even though the results are more comparable to the experiential values, it is far from satisfactory. Instead, the local field at $\mu1$ site needs to increase compared with the one at $\mu2$. Since the largest, site dependent, contributions to the local field is the dipolar field, the ordered moments need to be ordered more in plane where the $\mu1$ site is located. In fact, tiling the ordered moments from c-axis by $62^\circ$ (or $28^\circ$ from the a/b-plane) shifts the local field ratio to 2.17, comparable to experiment. The local field values for this kind of magnetic structure, where the orders moments are set to $6.5~\mu_{\rm B}$, are shown in table~\ref{table:Fre2}. In this scenario, the calculated local field is consistent with the experimental results. 

An ordered magnetic moment of $6.5~\mu_{\rm B}$ is significantly higher than the expected pure spin moment of $2.82~\mu_{\rm B} /V^{3+}$. Since magnetisation measurements determine an effective moment of about $2.6~\mu_{\rm B}$ per V$^{3+}$, it seems as the structural transition at 80~K changes the magnetic interactions such that orbital angular momentum becomes significant. V$^{3+}$ has the electron configuration 3d2 and the ordered moment for this kind of system is expected to be $\sqrt(L(L+1)+4S(S+1))=4.47~\mu_{\rm B}$, which is consistent with our findings. This is however somewhat surprising considering that V is in a octahedral environment and the orbital contributions can usually be neglected. In fact, the situation regarding the angular orbital momentum of V$^{3+}$ in VI$_3$ has been under a debate \cite{Yang2020, Nguyen2021}. With that said, our results represents clear experimental proof of unquenched orbital angular momentum of V$^{3+}$ in VI$_3$. 
This is consistent with electronic structure calculations \cite{Yang2020} suggesting unquenched orbital angular momentum, arising from the electronic configuration $a_{1g}'e^{'1}_{-}$.  



Finally, we wish to discuss the origin to the behaviour observed between $T_{\rm C} < T \leq 58$~K. It is noted that the behaviour observed here does not correspond to the reported anomalies at $T_1 = 54.5$~K and $T_2 =53$~K in Ref.~\onlinecite{Valenta2021}.  $T_1$ and $T_2$ were speculated to originate from surface degradation, a phenomenon which can be excluded in this study as the sample was mounted and sealed in He atmosphere. Instead, we shall note that the behaviour observed in this study resembles the ones typically observed in low dimensional systems. In these systems, a short range magnetic order is formed just above $T_{\rm C}$ and result in a slow exponential component ($A_{\rm S}$) which increase and a fast exponential component ($A_{\rm F}$) which decreases with increasing temperature. Such ordering has been observed and recently clarified in the layered sister compound CrCl$_3$ \cite{Forslund2021}, but also in NaNiO$_2$ \cite{Baker2005, Forslund_2020_Na}, LaCoP$_2$ \cite{Forslund_2021_La}, BaCo$_2$V$_2$O$_{8}$ \cite{Mansson2012}, $A_{n+2}$Co$_{n+1}$O$_{3n+3}$ ($A$=Ca, Sr, Ba) \cite{Sugiyama2005,Sugiyama2006} or $A_{n+2}$CoRh$_n$O$_{3n+3}$ ($A$=Ca, Sr) \cite{Sugiyama2008}. 

For a long range magnetic ordered case, the spin-spin correlation function ($C$) does not decay while a power law decay $C \propto r^{-d}$ is expected for a quasi long range order, $e.g.$ in Kosterlitz-Thouless transition \cite{Schollwock2008}. A system with an exponential decay on the other hand ($C \propto e^{-r/\theta}$) can be considered a short range ordered system with a correlation length $\theta$. These systems can be experientially observed in static thermodynamic measurements \cite{Baker2005, He2005, Kong2019, Yamauchi2011, McGuire2017}. Since bulk thermodynamic measurements has not observed any transition to be present around 58~K in VI$_3$ \cite{Tian2019, Gati2019, Marchandier2021}, we propose that the origin to the anomaly is hidden in the dynamical nature. Therefore, an exponential decay of the spin correlation function in time (instead of space) may be the cause of this behaviour in VI$_3$. To confirm the scenario, it is highly desirable to perform spin echo measurements to directly deduce the correlation decays. However, the elements V and I may prove neutron studies difficult. 


\section{\label{sec:conclusion}Conclusions}
The ferromagnetic (FM) VI$_3$ was measured as a function of temperature using muon spin relaxation ($\mu^+$SR). The appearance of muon spin precession frequencies for $T<T_{\rm C}=50.03(1)$~K confirms the formation of FM order in the whole bulk. By combining $\mu^+$SR and density functional theory (DFT) calculations, the expected internal magnetic field is calculated based on the established magnetic structure. These calculation suggest that the established magnetic structure is not accurate and the ordered magnetic moment is likely to be higher, around 6~$\mu_{\rm B}$. Moreover, the FM structure is most likely titled further away from the $c$-axis than previously suggested. These results represent experimental evidence of a strong orbital angular momentum contribution to the magnetism in this compound. This work highlights $\mu^+$SR as a viable choice in determining the magnetic structure of different magnetic compounds, including weakly ordered systems or systems containing elements not compatible with neutrons, e.g. Eu, V or I. Finally, a short range ordered phase was found just above $T_{\rm C}$, similar to other layered compounds. However, the SRO phase in VI$_3$ exhibits a temperature dependence unlike the others and additional detailed studies are desirable.

\begin{acknowledgments}

\end{acknowledgments}

\bibliography{Refs} 

\begin{thebibliography}{43}%
\makeatletter
\providecommand \@ifxundefined [1]{%
 \@ifx{#1\undefined}
}%
\providecommand \@ifnum [1]{%
 \ifnum #1\expandafter \@firstoftwo
 \else \expandafter \@secondoftwo
 \fi
}%
\providecommand \@ifx [1]{%
 \ifx #1\expandafter \@firstoftwo
 \else \expandafter \@secondoftwo
 \fi
}%
\providecommand \natexlab [1]{#1}%
\providecommand \enquote  [1]{``#1''}%
\providecommand \bibnamefont  [1]{#1}%
\providecommand \bibfnamefont [1]{#1}%
\providecommand \citenamefont [1]{#1}%
\providecommand \href@noop [0]{\@secondoftwo}%
\providecommand \href [0]{\begingroup \@sanitize@url \@href}%
\providecommand \@href[1]{\@@startlink{#1}\@@href}%
\providecommand \@@href[1]{\endgroup#1\@@endlink}%
\providecommand \@sanitize@url [0]{\catcode `\\12\catcode `\$12\catcode
  `\&12\catcode `\#12\catcode `\^12\catcode `\_12\catcode `\%12\relax}%
\providecommand \@@startlink[1]{}%
\providecommand \@@endlink[0]{}%
\providecommand \url  [0]{\begingroup\@sanitize@url \@url }%
\providecommand \@url [1]{\endgroup\@href {#1}{\urlprefix }}%
\providecommand \urlprefix  [0]{URL }%
\providecommand \Eprint [0]{\href }%
\providecommand \doibase [0]{http://dx.doi.org/}%
\providecommand \selectlanguage [0]{\@gobble}%
\providecommand \bibinfo  [0]{\@secondoftwo}%
\providecommand \bibfield  [0]{\@secondoftwo}%
\providecommand \translation [1]{[#1]}%
\providecommand \BibitemOpen [0]{}%
\providecommand \bibitemStop [0]{}%
\providecommand \bibitemNoStop [0]{.\EOS\space}%
\providecommand \EOS [0]{\spacefactor3000\relax}%
\providecommand \BibitemShut  [1]{\csname bibitem#1\endcsname}%
\let\auto@bib@innerbib\@empty
\bibitem [{\citenamefont {Ising}(1924)}]{Ising1924}%
  \BibitemOpen
  \bibfield  {author} {\bibinfo {author} {\bibfnamefont {E.}~\bibnamefont
  {Ising}},\ }\emph {\bibinfo {title} {Beitrag zur theorie des ferro-und
  paramagnetismus}},\ \href@noop {} {Ph.D. thesis},\ \bibinfo  {school} {Grefe
  \& Tiedemann} (\bibinfo {year} {1924})\BibitemShut {NoStop}%
\bibitem [{\citenamefont {Huang}\ \emph {et~al.}(2017)\citenamefont {Huang},
  \citenamefont {Clark}, \citenamefont {Navarro-Moratalla}, \citenamefont
  {Klein}, \citenamefont {Cheng}, \citenamefont {Seyler}, \citenamefont
  {Zhong}, \citenamefont {Schmidgall}, \citenamefont {McGuire}, \citenamefont
  {Cobden}, \citenamefont {Yao}, \citenamefont {Xiao}, \citenamefont
  {Jarillo-Herrero},\ and\ \citenamefont {Xu}}]{Huang2017}%
  \BibitemOpen
  \bibfield  {author} {\bibinfo {author} {\bibfnamefont {B.}~\bibnamefont
  {Huang}}, \bibinfo {author} {\bibfnamefont {G.}~\bibnamefont {Clark}},
  \bibinfo {author} {\bibfnamefont {E.}~\bibnamefont {Navarro-Moratalla}},
  \bibinfo {author} {\bibfnamefont {D.~R.}\ \bibnamefont {Klein}}, \bibinfo
  {author} {\bibfnamefont {R.}~\bibnamefont {Cheng}}, \bibinfo {author}
  {\bibfnamefont {K.~L.}\ \bibnamefont {Seyler}}, \bibinfo {author}
  {\bibfnamefont {D.}~\bibnamefont {Zhong}}, \bibinfo {author} {\bibfnamefont
  {E.}~\bibnamefont {Schmidgall}}, \bibinfo {author} {\bibfnamefont {M.~A.}\
  \bibnamefont {McGuire}}, \bibinfo {author} {\bibfnamefont {D.~H.}\
  \bibnamefont {Cobden}}, \bibinfo {author} {\bibfnamefont {W.}~\bibnamefont
  {Yao}}, \bibinfo {author} {\bibfnamefont {D.}~\bibnamefont {Xiao}}, \bibinfo
  {author} {\bibfnamefont {P.}~\bibnamefont {Jarillo-Herrero}}, \ and\ \bibinfo
  {author} {\bibfnamefont {X.}~\bibnamefont {Xu}},\ }\href {\doibase
  10.1038/nature22391} {\bibfield  {journal} {\bibinfo  {journal} {Nature}\
  }\textbf {\bibinfo {volume} {546}},\ \bibinfo {pages} {270} (\bibinfo {year}
  {2017})}\BibitemShut {NoStop}%
\bibitem [{\citenamefont {Matsukura}\ \emph {et~al.}(2015)\citenamefont
  {Matsukura}, \citenamefont {Tokura},\ and\ \citenamefont
  {Ohno}}]{Matsukura2015}%
  \BibitemOpen
  \bibfield  {author} {\bibinfo {author} {\bibfnamefont {F.}~\bibnamefont
  {Matsukura}}, \bibinfo {author} {\bibfnamefont {Y.}~\bibnamefont {Tokura}}, \
  and\ \bibinfo {author} {\bibfnamefont {H.}~\bibnamefont {Ohno}},\ }\href
  {\doibase 10.1038/nnano.2015.22} {\bibfield  {journal} {\bibinfo  {journal}
  {Nature Nanotechnology}\ }\textbf {\bibinfo {volume} {10}},\ \bibinfo {pages}
  {209} (\bibinfo {year} {2015})}\BibitemShut {NoStop}%
\bibitem [{\citenamefont {Jiang}\ \emph {et~al.}(2018)\citenamefont {Jiang},
  \citenamefont {Li}, \citenamefont {Wang}, \citenamefont {Mak},\ and\
  \citenamefont {Shan}}]{Jiang2018}%
  \BibitemOpen
  \bibfield  {author} {\bibinfo {author} {\bibfnamefont {S.}~\bibnamefont
  {Jiang}}, \bibinfo {author} {\bibfnamefont {L.}~\bibnamefont {Li}}, \bibinfo
  {author} {\bibfnamefont {Z.}~\bibnamefont {Wang}}, \bibinfo {author}
  {\bibfnamefont {K.~F.}\ \bibnamefont {Mak}}, \ and\ \bibinfo {author}
  {\bibfnamefont {J.}~\bibnamefont {Shan}},\ }\href {\doibase
  10.1038/s41565-018-0135-x} {\bibfield  {journal} {\bibinfo  {journal} {Nature
  Nanotechnology}\ }\textbf {\bibinfo {volume} {13}},\ \bibinfo {pages} {549}
  (\bibinfo {year} {2018})}\BibitemShut {NoStop}%
\bibitem [{\citenamefont {Kong}\ \emph {et~al.}(2019)\citenamefont {Kong},
  \citenamefont {Stolze}, \citenamefont {Timmons}, \citenamefont {Tao},
  \citenamefont {Ni}, \citenamefont {Guo}, \citenamefont {Yang}, \citenamefont
  {Prozorov},\ and\ \citenamefont {Cava}}]{Kong2019}%
  \BibitemOpen
  \bibfield  {author} {\bibinfo {author} {\bibfnamefont {T.}~\bibnamefont
  {Kong}}, \bibinfo {author} {\bibfnamefont {K.}~\bibnamefont {Stolze}},
  \bibinfo {author} {\bibfnamefont {E.~I.}\ \bibnamefont {Timmons}}, \bibinfo
  {author} {\bibfnamefont {J.}~\bibnamefont {Tao}}, \bibinfo {author}
  {\bibfnamefont {D.}~\bibnamefont {Ni}}, \bibinfo {author} {\bibfnamefont
  {S.}~\bibnamefont {Guo}}, \bibinfo {author} {\bibfnamefont {Z.}~\bibnamefont
  {Yang}}, \bibinfo {author} {\bibfnamefont {R.}~\bibnamefont {Prozorov}}, \
  and\ \bibinfo {author} {\bibfnamefont {R.~J.}\ \bibnamefont {Cava}},\ }\href
  {\doibase https://doi.org/10.1002/adma.201808074} {\bibfield  {journal}
  {\bibinfo  {journal} {Advanced Materials}\ }\textbf {\bibinfo {volume}
  {31}},\ \bibinfo {pages} {1808074} (\bibinfo {year} {2019})},\ \Eprint
  {http://arxiv.org/abs/https://onlinelibrary.wiley.com/doi/pdf/10.1002/adma.201808074}
  {https://onlinelibrary.wiley.com/doi/pdf/10.1002/adma.201808074} \BibitemShut
  {NoStop}%
\bibitem [{\citenamefont {Tian}\ \emph {et~al.}(2019)\citenamefont {Tian},
  \citenamefont {Zhang}, \citenamefont {Li}, \citenamefont {Ying},
  \citenamefont {Li}, \citenamefont {Zhang}, \citenamefont {Liu},\ and\
  \citenamefont {Lei}}]{Tian2019}%
  \BibitemOpen
  \bibfield  {author} {\bibinfo {author} {\bibfnamefont {S.}~\bibnamefont
  {Tian}}, \bibinfo {author} {\bibfnamefont {J.-F.}\ \bibnamefont {Zhang}},
  \bibinfo {author} {\bibfnamefont {C.}~\bibnamefont {Li}}, \bibinfo {author}
  {\bibfnamefont {T.}~\bibnamefont {Ying}}, \bibinfo {author} {\bibfnamefont
  {S.}~\bibnamefont {Li}}, \bibinfo {author} {\bibfnamefont {X.}~\bibnamefont
  {Zhang}}, \bibinfo {author} {\bibfnamefont {K.}~\bibnamefont {Liu}}, \ and\
  \bibinfo {author} {\bibfnamefont {H.}~\bibnamefont {Lei}},\ }\href {\doibase
  10.1021/jacs.8b13584} {\bibfield  {journal} {\bibinfo  {journal} {Journal of
  the American Chemical Society}\ }\textbf {\bibinfo {volume} {141}},\ \bibinfo
  {pages} {5326} (\bibinfo {year} {2019})}\BibitemShut {NoStop}%
\bibitem [{\citenamefont {Son}\ \emph {et~al.}(2019)\citenamefont {Son},
  \citenamefont {Coak}, \citenamefont {Lee}, \citenamefont {Kim}, \citenamefont
  {Kim}, \citenamefont {Hamidov}, \citenamefont {Cho}, \citenamefont {Liu},
  \citenamefont {Jarvis}, \citenamefont {Brown}, \citenamefont {Kim},
  \citenamefont {Park}, \citenamefont {Khomskii}, \citenamefont {Saxena},\ and\
  \citenamefont {Park}}]{Son2019}%
  \BibitemOpen
  \bibfield  {author} {\bibinfo {author} {\bibfnamefont {S.}~\bibnamefont
  {Son}}, \bibinfo {author} {\bibfnamefont {M.~J.}\ \bibnamefont {Coak}},
  \bibinfo {author} {\bibfnamefont {N.}~\bibnamefont {Lee}}, \bibinfo {author}
  {\bibfnamefont {J.}~\bibnamefont {Kim}}, \bibinfo {author} {\bibfnamefont
  {T.~Y.}\ \bibnamefont {Kim}}, \bibinfo {author} {\bibfnamefont
  {H.}~\bibnamefont {Hamidov}}, \bibinfo {author} {\bibfnamefont
  {H.}~\bibnamefont {Cho}}, \bibinfo {author} {\bibfnamefont {C.}~\bibnamefont
  {Liu}}, \bibinfo {author} {\bibfnamefont {D.~M.}\ \bibnamefont {Jarvis}},
  \bibinfo {author} {\bibfnamefont {P.~A.~C.}\ \bibnamefont {Brown}}, \bibinfo
  {author} {\bibfnamefont {J.~H.}\ \bibnamefont {Kim}}, \bibinfo {author}
  {\bibfnamefont {C.-H.}\ \bibnamefont {Park}}, \bibinfo {author}
  {\bibfnamefont {D.~I.}\ \bibnamefont {Khomskii}}, \bibinfo {author}
  {\bibfnamefont {S.~S.}\ \bibnamefont {Saxena}}, \ and\ \bibinfo {author}
  {\bibfnamefont {J.-G.}\ \bibnamefont {Park}},\ }\href {\doibase
  10.1103/PhysRevB.99.041402} {\bibfield  {journal} {\bibinfo  {journal} {Phys.
  Rev. B}\ }\textbf {\bibinfo {volume} {99}},\ \bibinfo {pages} {041402}
  (\bibinfo {year} {2019})}\BibitemShut {NoStop}%
\bibitem [{\citenamefont {Dole\ifmmode~\check{z}\else \v{z}\fi{}al}\ \emph
  {et~al.}(2019)\citenamefont {Dole\ifmmode~\check{z}\else \v{z}\fi{}al},
  \citenamefont {Kratochv\'{\i}lov\'a}, \citenamefont {Hol\'y}, \citenamefont
  {\ifmmode~\check{C}\else \v{C}\fi{}erm\'ak}, \citenamefont {Sechovsk\'y},
  \citenamefont {Du\ifmmode~\check{s}\else \v{s}\fi{}ek}, \citenamefont
  {M\'{\i}\ifmmode~\check{s}\else \v{s}\fi{}ek}, \citenamefont {Chakraborty},
  \citenamefont {Noda}, \citenamefont {Son},\ and\ \citenamefont
  {Park}}]{Doleifmmode2019}%
  \BibitemOpen
  \bibfield  {author} {\bibinfo {author} {\bibfnamefont {P.}~\bibnamefont
  {Dole\ifmmode~\check{z}\else \v{z}\fi{}al}}, \bibinfo {author} {\bibfnamefont
  {M.}~\bibnamefont {Kratochv\'{\i}lov\'a}}, \bibinfo {author} {\bibfnamefont
  {V.}~\bibnamefont {Hol\'y}}, \bibinfo {author} {\bibfnamefont
  {P.}~\bibnamefont {\ifmmode~\check{C}\else \v{C}\fi{}erm\'ak}}, \bibinfo
  {author} {\bibfnamefont {V.}~\bibnamefont {Sechovsk\'y}}, \bibinfo {author}
  {\bibfnamefont {M.}~\bibnamefont {Du\ifmmode~\check{s}\else \v{s}\fi{}ek}},
  \bibinfo {author} {\bibfnamefont {M.}~\bibnamefont
  {M\'{\i}\ifmmode~\check{s}\else \v{s}\fi{}ek}}, \bibinfo {author}
  {\bibfnamefont {T.}~\bibnamefont {Chakraborty}}, \bibinfo {author}
  {\bibfnamefont {Y.}~\bibnamefont {Noda}}, \bibinfo {author} {\bibfnamefont
  {S.}~\bibnamefont {Son}}, \ and\ \bibinfo {author} {\bibfnamefont {J.-G.}\
  \bibnamefont {Park}},\ }\href {\doibase 10.1103/PhysRevMaterials.3.121401}
  {\bibfield  {journal} {\bibinfo  {journal} {Phys. Rev. Materials}\ }\textbf
  {\bibinfo {volume} {3}},\ \bibinfo {pages} {121401} (\bibinfo {year}
  {2019})}\BibitemShut {NoStop}%
\bibitem [{\citenamefont {Marchandier}\ \emph {et~al.}(2021)\citenamefont
  {Marchandier}, \citenamefont {Dubouis}, \citenamefont {Fauth}, \citenamefont
  {Avdeev}, \citenamefont {Grimaud}, \citenamefont {Tarascon},\ and\
  \citenamefont {Rousse}}]{Marchandier2021}%
  \BibitemOpen
  \bibfield  {author} {\bibinfo {author} {\bibfnamefont {T.}~\bibnamefont
  {Marchandier}}, \bibinfo {author} {\bibfnamefont {N.}~\bibnamefont
  {Dubouis}}, \bibinfo {author} {\bibfnamefont {F.~m.~c.}\ \bibnamefont
  {Fauth}}, \bibinfo {author} {\bibfnamefont {M.}~\bibnamefont {Avdeev}},
  \bibinfo {author} {\bibfnamefont {A.}~\bibnamefont {Grimaud}}, \bibinfo
  {author} {\bibfnamefont {J.-M.}\ \bibnamefont {Tarascon}}, \ and\ \bibinfo
  {author} {\bibfnamefont {G.}~\bibnamefont {Rousse}},\ }\href {\doibase
  10.1103/PhysRevB.104.014105} {\bibfield  {journal} {\bibinfo  {journal}
  {Phys. Rev. B}\ }\textbf {\bibinfo {volume} {104}},\ \bibinfo {pages}
  {014105} (\bibinfo {year} {2021})}\BibitemShut {NoStop}%
\bibitem [{\citenamefont {Valenta}\ \emph {et~al.}(2021)\citenamefont
  {Valenta}, \citenamefont {Kratochv\'{\i}lov\'a}, \citenamefont
  {M\'{\i}\ifmmode~\check{s}\else \v{s}\fi{}ek}, \citenamefont {Carva},
  \citenamefont {Ka\ifmmode~\check{s}\else \v{s}\fi{}til}, \citenamefont
  {Dole\ifmmode~\check{z}\else \v{z}\fi{}al}, \citenamefont {Opletal},
  \citenamefont {\ifmmode~\check{C}\else \v{C}\fi{}erm\'ak}, \citenamefont
  {Proschek}, \citenamefont {Uhl\'{\i}\ifmmode~\check{r}\else \v{r}\fi{}ov\'a},
  \citenamefont {Prchal}, \citenamefont {Coak}, \citenamefont {Son},
  \citenamefont {Park},\ and\ \citenamefont {Sechovsk\'y}}]{Valenta2021}%
  \BibitemOpen
  \bibfield  {author} {\bibinfo {author} {\bibfnamefont {J.}~\bibnamefont
  {Valenta}}, \bibinfo {author} {\bibfnamefont {M.}~\bibnamefont
  {Kratochv\'{\i}lov\'a}}, \bibinfo {author} {\bibfnamefont {M.}~\bibnamefont
  {M\'{\i}\ifmmode~\check{s}\else \v{s}\fi{}ek}}, \bibinfo {author}
  {\bibfnamefont {K.}~\bibnamefont {Carva}}, \bibinfo {author} {\bibfnamefont
  {J.}~\bibnamefont {Ka\ifmmode~\check{s}\else \v{s}\fi{}til}}, \bibinfo
  {author} {\bibfnamefont {P.}~\bibnamefont {Dole\ifmmode~\check{z}\else
  \v{z}\fi{}al}}, \bibinfo {author} {\bibfnamefont {P.}~\bibnamefont
  {Opletal}}, \bibinfo {author} {\bibfnamefont {P.}~\bibnamefont
  {\ifmmode~\check{C}\else \v{C}\fi{}erm\'ak}}, \bibinfo {author}
  {\bibfnamefont {P.}~\bibnamefont {Proschek}}, \bibinfo {author}
  {\bibfnamefont {K.}~\bibnamefont {Uhl\'{\i}\ifmmode~\check{r}\else
  \v{r}\fi{}ov\'a}}, \bibinfo {author} {\bibfnamefont {J.}~\bibnamefont
  {Prchal}}, \bibinfo {author} {\bibfnamefont {M.~J.}\ \bibnamefont {Coak}},
  \bibinfo {author} {\bibfnamefont {S.}~\bibnamefont {Son}}, \bibinfo {author}
  {\bibfnamefont {J.-G.}\ \bibnamefont {Park}}, \ and\ \bibinfo {author}
  {\bibfnamefont {V.}~\bibnamefont {Sechovsk\'y}},\ }\href {\doibase
  10.1103/PhysRevB.103.054424} {\bibfield  {journal} {\bibinfo  {journal}
  {Phys. Rev. B}\ }\textbf {\bibinfo {volume} {103}},\ \bibinfo {pages}
  {054424} (\bibinfo {year} {2021})}\BibitemShut {NoStop}%
\bibitem [{\citenamefont {Gati}\ \emph {et~al.}(2019)\citenamefont {Gati},
  \citenamefont {Inagaki}, \citenamefont {Kong}, \citenamefont {Cava},
  \citenamefont {Furukawa}, \citenamefont {Canfield},\ and\ \citenamefont
  {Bud'ko}}]{Gati2019}%
  \BibitemOpen
  \bibfield  {author} {\bibinfo {author} {\bibfnamefont {E.}~\bibnamefont
  {Gati}}, \bibinfo {author} {\bibfnamefont {Y.}~\bibnamefont {Inagaki}},
  \bibinfo {author} {\bibfnamefont {T.}~\bibnamefont {Kong}}, \bibinfo {author}
  {\bibfnamefont {R.~J.}\ \bibnamefont {Cava}}, \bibinfo {author}
  {\bibfnamefont {Y.}~\bibnamefont {Furukawa}}, \bibinfo {author}
  {\bibfnamefont {P.~C.}\ \bibnamefont {Canfield}}, \ and\ \bibinfo {author}
  {\bibfnamefont {S.~L.}\ \bibnamefont {Bud'ko}},\ }\href {\doibase
  10.1103/PhysRevB.100.094408} {\bibfield  {journal} {\bibinfo  {journal}
  {Phys. Rev. B}\ }\textbf {\bibinfo {volume} {100}},\ \bibinfo {pages}
  {094408} (\bibinfo {year} {2019})}\BibitemShut {NoStop}%
\bibitem [{\citenamefont {Chen}\ \emph {et~al.}(2018)\citenamefont {Chen},
  \citenamefont {Chung}, \citenamefont {Gao}, \citenamefont {Chen},
  \citenamefont {Stone}, \citenamefont {Kolesnikov}, \citenamefont {Huang},\
  and\ \citenamefont {Dai}}]{Chen2018}%
  \BibitemOpen
  \bibfield  {author} {\bibinfo {author} {\bibfnamefont {L.}~\bibnamefont
  {Chen}}, \bibinfo {author} {\bibfnamefont {J.-H.}\ \bibnamefont {Chung}},
  \bibinfo {author} {\bibfnamefont {B.}~\bibnamefont {Gao}}, \bibinfo {author}
  {\bibfnamefont {T.}~\bibnamefont {Chen}}, \bibinfo {author} {\bibfnamefont
  {M.~B.}\ \bibnamefont {Stone}}, \bibinfo {author} {\bibfnamefont {A.~I.}\
  \bibnamefont {Kolesnikov}}, \bibinfo {author} {\bibfnamefont
  {Q.}~\bibnamefont {Huang}}, \ and\ \bibinfo {author} {\bibfnamefont
  {P.}~\bibnamefont {Dai}},\ }\href {\doibase 10.1103/PhysRevX.8.041028}
  {\bibfield  {journal} {\bibinfo  {journal} {Phys. Rev. X}\ }\textbf {\bibinfo
  {volume} {8}},\ \bibinfo {pages} {041028} (\bibinfo {year}
  {2018})}\BibitemShut {NoStop}%
\bibitem [{\citenamefont {Chen}\ \emph {et~al.}(2020)\citenamefont {Chen},
  \citenamefont {Chung}, \citenamefont {Chen}, \citenamefont {Duan},
  \citenamefont {Schneidewind}, \citenamefont {Radelytskyi}, \citenamefont
  {Voneshen}, \citenamefont {Ewings}, \citenamefont {Stone}, \citenamefont
  {Kolesnikov}, \citenamefont {Winn}, \citenamefont {Chi}, \citenamefont
  {Mole}, \citenamefont {Yu}, \citenamefont {Gao},\ and\ \citenamefont
  {Dai}}]{Chen2020}%
  \BibitemOpen
  \bibfield  {author} {\bibinfo {author} {\bibfnamefont {L.}~\bibnamefont
  {Chen}}, \bibinfo {author} {\bibfnamefont {J.-H.}\ \bibnamefont {Chung}},
  \bibinfo {author} {\bibfnamefont {T.}~\bibnamefont {Chen}}, \bibinfo {author}
  {\bibfnamefont {C.}~\bibnamefont {Duan}}, \bibinfo {author} {\bibfnamefont
  {A.}~\bibnamefont {Schneidewind}}, \bibinfo {author} {\bibfnamefont
  {I.}~\bibnamefont {Radelytskyi}}, \bibinfo {author} {\bibfnamefont {D.~J.}\
  \bibnamefont {Voneshen}}, \bibinfo {author} {\bibfnamefont {R.~A.}\
  \bibnamefont {Ewings}}, \bibinfo {author} {\bibfnamefont {M.~B.}\
  \bibnamefont {Stone}}, \bibinfo {author} {\bibfnamefont {A.~I.}\ \bibnamefont
  {Kolesnikov}}, \bibinfo {author} {\bibfnamefont {B.}~\bibnamefont {Winn}},
  \bibinfo {author} {\bibfnamefont {S.}~\bibnamefont {Chi}}, \bibinfo {author}
  {\bibfnamefont {R.~A.}\ \bibnamefont {Mole}}, \bibinfo {author}
  {\bibfnamefont {D.~H.}\ \bibnamefont {Yu}}, \bibinfo {author} {\bibfnamefont
  {B.}~\bibnamefont {Gao}}, \ and\ \bibinfo {author} {\bibfnamefont
  {P.}~\bibnamefont {Dai}},\ }\href {\doibase 10.1103/PhysRevB.101.134418}
  {\bibfield  {journal} {\bibinfo  {journal} {Phys. Rev. B}\ }\textbf {\bibinfo
  {volume} {101}},\ \bibinfo {pages} {134418} (\bibinfo {year}
  {2020})}\BibitemShut {NoStop}%
\bibitem [{\citenamefont {Cai}\ \emph {et~al.}(2021)\citenamefont {Cai},
  \citenamefont {Bao}, \citenamefont {Gu}, \citenamefont {Gao}, \citenamefont
  {Ma}, \citenamefont {Shangguan}, \citenamefont {Si}, \citenamefont {Dong},
  \citenamefont {Wang}, \citenamefont {Wu}, \citenamefont {Lin}, \citenamefont
  {Wang}, \citenamefont {Ran}, \citenamefont {Li}, \citenamefont {Adroja},
  \citenamefont {Xi}, \citenamefont {Yu}, \citenamefont {Wu}, \citenamefont
  {Li},\ and\ \citenamefont {Wen}}]{Cai2021}%
  \BibitemOpen
  \bibfield  {author} {\bibinfo {author} {\bibfnamefont {Z.}~\bibnamefont
  {Cai}}, \bibinfo {author} {\bibfnamefont {S.}~\bibnamefont {Bao}}, \bibinfo
  {author} {\bibfnamefont {Z.-L.}\ \bibnamefont {Gu}}, \bibinfo {author}
  {\bibfnamefont {Y.-P.}\ \bibnamefont {Gao}}, \bibinfo {author} {\bibfnamefont
  {Z.}~\bibnamefont {Ma}}, \bibinfo {author} {\bibfnamefont {Y.}~\bibnamefont
  {Shangguan}}, \bibinfo {author} {\bibfnamefont {W.}~\bibnamefont {Si}},
  \bibinfo {author} {\bibfnamefont {Z.-Y.}\ \bibnamefont {Dong}}, \bibinfo
  {author} {\bibfnamefont {W.}~\bibnamefont {Wang}}, \bibinfo {author}
  {\bibfnamefont {Y.}~\bibnamefont {Wu}}, \bibinfo {author} {\bibfnamefont
  {D.}~\bibnamefont {Lin}}, \bibinfo {author} {\bibfnamefont {J.}~\bibnamefont
  {Wang}}, \bibinfo {author} {\bibfnamefont {K.}~\bibnamefont {Ran}}, \bibinfo
  {author} {\bibfnamefont {S.}~\bibnamefont {Li}}, \bibinfo {author}
  {\bibfnamefont {D.}~\bibnamefont {Adroja}}, \bibinfo {author} {\bibfnamefont
  {X.}~\bibnamefont {Xi}}, \bibinfo {author} {\bibfnamefont {S.-L.}\
  \bibnamefont {Yu}}, \bibinfo {author} {\bibfnamefont {X.}~\bibnamefont {Wu}},
  \bibinfo {author} {\bibfnamefont {J.-X.}\ \bibnamefont {Li}}, \ and\ \bibinfo
  {author} {\bibfnamefont {J.}~\bibnamefont {Wen}},\ }\href {\doibase
  10.1103/PhysRevB.104.L020402} {\bibfield  {journal} {\bibinfo  {journal}
  {Phys. Rev. B}\ }\textbf {\bibinfo {volume} {104}},\ \bibinfo {pages}
  {L020402} (\bibinfo {year} {2021})}\BibitemShut {NoStop}%
\bibitem [{\citenamefont {Chen}\ \emph {et~al.}(2021)\citenamefont {Chen},
  \citenamefont {Stone}, \citenamefont {Kolesnikov}, \citenamefont {Winn},
  \citenamefont {Shon}, \citenamefont {Dai},\ and\ \citenamefont
  {Chung}}]{Chen2022}%
  \BibitemOpen
  \bibfield  {author} {\bibinfo {author} {\bibfnamefont {L.}~\bibnamefont
  {Chen}}, \bibinfo {author} {\bibfnamefont {M.~B.}\ \bibnamefont {Stone}},
  \bibinfo {author} {\bibfnamefont {A.~I.}\ \bibnamefont {Kolesnikov}},
  \bibinfo {author} {\bibfnamefont {B.}~\bibnamefont {Winn}}, \bibinfo {author}
  {\bibfnamefont {W.}~\bibnamefont {Shon}}, \bibinfo {author} {\bibfnamefont
  {P.}~\bibnamefont {Dai}}, \ and\ \bibinfo {author} {\bibfnamefont {J.-H.}\
  \bibnamefont {Chung}},\ }\href {\doibase 10.1088/2053-1583/ac2e7a} {\bibfield
   {journal} {\bibinfo  {journal} {2D Materials}\ }\textbf {\bibinfo {volume}
  {9}},\ \bibinfo {pages} {015006} (\bibinfo {year} {2021})}\BibitemShut
  {NoStop}%
\bibitem [{\citenamefont {Do}\ \emph {et~al.}(2022)\citenamefont {Do},
  \citenamefont {Paddison}, \citenamefont {Sala}, \citenamefont {Williams},
  \citenamefont {Kaneko}, \citenamefont {Kuwahara}, \citenamefont {May},
  \citenamefont {Yan}, \citenamefont {McGuire}, \citenamefont {Stone},
  \citenamefont {Lumsden},\ and\ \citenamefont {Christianson}}]{Do2022}%
  \BibitemOpen
  \bibfield  {author} {\bibinfo {author} {\bibfnamefont {S.-H.}\ \bibnamefont
  {Do}}, \bibinfo {author} {\bibfnamefont {J.~A.~M.}\ \bibnamefont {Paddison}},
  \bibinfo {author} {\bibfnamefont {G.}~\bibnamefont {Sala}}, \bibinfo {author}
  {\bibfnamefont {T.~J.}\ \bibnamefont {Williams}}, \bibinfo {author}
  {\bibfnamefont {K.}~\bibnamefont {Kaneko}}, \bibinfo {author} {\bibfnamefont
  {K.}~\bibnamefont {Kuwahara}}, \bibinfo {author} {\bibfnamefont {A.~F.}\
  \bibnamefont {May}}, \bibinfo {author} {\bibfnamefont {J.}~\bibnamefont
  {Yan}}, \bibinfo {author} {\bibfnamefont {M.~A.}\ \bibnamefont {McGuire}},
  \bibinfo {author} {\bibfnamefont {M.~B.}\ \bibnamefont {Stone}}, \bibinfo
  {author} {\bibfnamefont {M.~D.}\ \bibnamefont {Lumsden}}, \ and\ \bibinfo
  {author} {\bibfnamefont {A.~D.}\ \bibnamefont {Christianson}},\ }\href
  {\doibase 10.1103/physrevb.106.l060408} {\bibfield  {journal} {\bibinfo
  {journal} {Physical Review B}\ }\textbf {\bibinfo {volume} {106}} (\bibinfo
  {year} {2022}),\ 10.1103/physrevb.106.l060408}\BibitemShut {NoStop}%
\bibitem [{\citenamefont {Forslund}\ \emph
  {et~al.}(2021{\natexlab{a}})\citenamefont {Forslund}, \citenamefont
  {Papadopoulos}, \citenamefont {Nocerino}, \citenamefont {Di~Berardino},
  \citenamefont {Wang}, \citenamefont {Sugiyama}, \citenamefont {Andreica},
  \citenamefont {Vasiliev}, \citenamefont {Abdel-Hafiez}, \citenamefont
  {Månsson},\ and\ \citenamefont {Sassa}}]{Forslund2021}%
  \BibitemOpen
  \bibfield  {author} {\bibinfo {author} {\bibfnamefont {O.~K.}\ \bibnamefont
  {Forslund}}, \bibinfo {author} {\bibfnamefont {K.}~\bibnamefont
  {Papadopoulos}}, \bibinfo {author} {\bibfnamefont {E.}~\bibnamefont
  {Nocerino}}, \bibinfo {author} {\bibfnamefont {G.}~\bibnamefont
  {Di~Berardino}}, \bibinfo {author} {\bibfnamefont {C.}~\bibnamefont {Wang}},
  \bibinfo {author} {\bibfnamefont {J.}~\bibnamefont {Sugiyama}}, \bibinfo
  {author} {\bibfnamefont {D.}~\bibnamefont {Andreica}}, \bibinfo {author}
  {\bibfnamefont {A.~N.}\ \bibnamefont {Vasiliev}}, \bibinfo {author}
  {\bibfnamefont {M.}~\bibnamefont {Abdel-Hafiez}}, \bibinfo {author}
  {\bibfnamefont {M.}~\bibnamefont {Månsson}}, \ and\ \bibinfo {author}
  {\bibfnamefont {Y.}~\bibnamefont {Sassa}},\ }\href {\doibase
  10.48550/ARXIV.2111.06246} {\enquote {\bibinfo {title} {Spin dynamics in the
  van der waals magnet crcl$_3$},}\ } (\bibinfo {year}
  {2021}{\natexlab{a}})\BibitemShut {NoStop}%
\bibitem [{\citenamefont {Hao}\ \emph {et~al.}(2021)\citenamefont {Hao},
  \citenamefont {Gu}, \citenamefont {Gu}, \citenamefont {Feng}, \citenamefont
  {Cao}, \citenamefont {Chi}, \citenamefont {Wu},\ and\ \citenamefont
  {Zhao}}]{Hao2021}%
  \BibitemOpen
  \bibfield  {author} {\bibinfo {author} {\bibfnamefont {Y.}~\bibnamefont
  {Hao}}, \bibinfo {author} {\bibfnamefont {Y.}~\bibnamefont {Gu}}, \bibinfo
  {author} {\bibfnamefont {Y.}~\bibnamefont {Gu}}, \bibinfo {author}
  {\bibfnamefont {E.}~\bibnamefont {Feng}}, \bibinfo {author} {\bibfnamefont
  {H.}~\bibnamefont {Cao}}, \bibinfo {author} {\bibfnamefont {S.}~\bibnamefont
  {Chi}}, \bibinfo {author} {\bibfnamefont {H.}~\bibnamefont {Wu}}, \ and\
  \bibinfo {author} {\bibfnamefont {J.}~\bibnamefont {Zhao}},\ }\href {\doibase
  10.1088/0256-307x/38/9/096101} {\bibfield  {journal} {\bibinfo  {journal}
  {Chinese Physics Letters}\ }\textbf {\bibinfo {volume} {38}},\ \bibinfo
  {pages} {096101} (\bibinfo {year} {2021})}\BibitemShut {NoStop}%
\bibitem [{\citenamefont {{A. Suter and B.~M. Wojek}}(2012)}]{musrfit}%
  \BibitemOpen
  \bibfield  {author} {\bibinfo {author} {\bibnamefont {{A. Suter and B.~M.
  Wojek}}},\ }\href@noop {} {\bibfield  {journal} {\bibinfo  {journal} {Phys.
  Proc.}\ }\textbf {\bibinfo {volume} {30}},\ \bibinfo {pages} {69} (\bibinfo
  {year} {2012})}\BibitemShut {NoStop}%
\bibitem [{\citenamefont {Giannozzi}\ \emph {et~al.}(2009)\citenamefont
  {Giannozzi}, \citenamefont {Baroni}, \citenamefont {Bonini}, \citenamefont
  {Calandra}, \citenamefont {Car}, \citenamefont {Cavazzoni}, \citenamefont
  {Ceresoli}, \citenamefont {Chiarotti}, \citenamefont {Cococcioni},
  \citenamefont {Dabo}, \citenamefont {{Dal Corso}}, \citenamefont
  {de~Gironcoli}, \citenamefont {Fabris}, \citenamefont {Fratesi},
  \citenamefont {Gebauer}, \citenamefont {Gerstmann}, \citenamefont
  {Gougoussis}, \citenamefont {Kokalj}, \citenamefont {Lazzeri}, \citenamefont
  {Martin-Samos}, \citenamefont {Marzari}, \citenamefont {Mauri}, \citenamefont
  {Mazzarello}, \citenamefont {Paolini}, \citenamefont {Pasquarello},
  \citenamefont {Paulatto}, \citenamefont {Sbraccia}, \citenamefont {Scandolo},
  \citenamefont {Sclauzero}, \citenamefont {Seitsonen}, \citenamefont
  {Smogunov}, \citenamefont {Umari},\ and\ \citenamefont
  {Wentzcovitch}}]{QE-2009}%
  \BibitemOpen
  \bibfield  {author} {\bibinfo {author} {\bibfnamefont {P.}~\bibnamefont
  {Giannozzi}}, \bibinfo {author} {\bibfnamefont {S.}~\bibnamefont {Baroni}},
  \bibinfo {author} {\bibfnamefont {N.}~\bibnamefont {Bonini}}, \bibinfo
  {author} {\bibfnamefont {M.}~\bibnamefont {Calandra}}, \bibinfo {author}
  {\bibfnamefont {R.}~\bibnamefont {Car}}, \bibinfo {author} {\bibfnamefont
  {C.}~\bibnamefont {Cavazzoni}}, \bibinfo {author} {\bibfnamefont
  {D.}~\bibnamefont {Ceresoli}}, \bibinfo {author} {\bibfnamefont {G.~L.}\
  \bibnamefont {Chiarotti}}, \bibinfo {author} {\bibfnamefont {M.}~\bibnamefont
  {Cococcioni}}, \bibinfo {author} {\bibfnamefont {I.}~\bibnamefont {Dabo}},
  \bibinfo {author} {\bibfnamefont {A.}~\bibnamefont {{Dal Corso}}}, \bibinfo
  {author} {\bibfnamefont {S.}~\bibnamefont {de~Gironcoli}}, \bibinfo {author}
  {\bibfnamefont {S.}~\bibnamefont {Fabris}}, \bibinfo {author} {\bibfnamefont
  {G.}~\bibnamefont {Fratesi}}, \bibinfo {author} {\bibfnamefont
  {R.}~\bibnamefont {Gebauer}}, \bibinfo {author} {\bibfnamefont
  {U.}~\bibnamefont {Gerstmann}}, \bibinfo {author} {\bibfnamefont
  {C.}~\bibnamefont {Gougoussis}}, \bibinfo {author} {\bibfnamefont
  {A.}~\bibnamefont {Kokalj}}, \bibinfo {author} {\bibfnamefont
  {M.}~\bibnamefont {Lazzeri}}, \bibinfo {author} {\bibfnamefont
  {L.}~\bibnamefont {Martin-Samos}}, \bibinfo {author} {\bibfnamefont
  {N.}~\bibnamefont {Marzari}}, \bibinfo {author} {\bibfnamefont
  {F.}~\bibnamefont {Mauri}}, \bibinfo {author} {\bibfnamefont
  {R.}~\bibnamefont {Mazzarello}}, \bibinfo {author} {\bibfnamefont
  {S.}~\bibnamefont {Paolini}}, \bibinfo {author} {\bibfnamefont
  {A.}~\bibnamefont {Pasquarello}}, \bibinfo {author} {\bibfnamefont
  {L.}~\bibnamefont {Paulatto}}, \bibinfo {author} {\bibfnamefont
  {C.}~\bibnamefont {Sbraccia}}, \bibinfo {author} {\bibfnamefont
  {S.}~\bibnamefont {Scandolo}}, \bibinfo {author} {\bibfnamefont
  {G.}~\bibnamefont {Sclauzero}}, \bibinfo {author} {\bibfnamefont {A.~P.}\
  \bibnamefont {Seitsonen}}, \bibinfo {author} {\bibfnamefont {A.}~\bibnamefont
  {Smogunov}}, \bibinfo {author} {\bibfnamefont {P.}~\bibnamefont {Umari}}, \
  and\ \bibinfo {author} {\bibfnamefont {R.~M.}\ \bibnamefont {Wentzcovitch}},\
  }\href {http://www.quantum-espresso.org} {\bibfield  {journal} {\bibinfo
  {journal} {Journal of Physics: Condensed Matter}\ }\textbf {\bibinfo {volume}
  {21}},\ \bibinfo {pages} {395502 (19pp)} (\bibinfo {year}
  {2009})}\BibitemShut {NoStop}%
\bibitem [{\citenamefont {Giannozzi}\ \emph {et~al.}(2017)\citenamefont
  {Giannozzi}, \citenamefont {Andreussi}, \citenamefont {Brumme}, \citenamefont
  {Bunau}, \citenamefont {Nardelli}, \citenamefont {Calandra}, \citenamefont
  {Car}, \citenamefont {Cavazzoni}, \citenamefont {Ceresoli}, \citenamefont
  {Cococcioni}, \citenamefont {Colonna}, \citenamefont {Carnimeo},
  \citenamefont {Corso}, \citenamefont {de~Gironcoli}, \citenamefont {Delugas},
  \citenamefont {Jr}, \citenamefont {Ferretti}, \citenamefont {Floris},
  \citenamefont {Fratesi}, \citenamefont {Fugallo}, \citenamefont {Gebauer},
  \citenamefont {Gerstmann}, \citenamefont {Giustino}, \citenamefont {Gorni},
  \citenamefont {Jia}, \citenamefont {Kawamura}, \citenamefont {Ko},
  \citenamefont {Kokalj}, \citenamefont {Küçükbenli}, \citenamefont
  {Lazzeri}, \citenamefont {Marsili}, \citenamefont {Marzari}, \citenamefont
  {Mauri}, \citenamefont {Nguyen}, \citenamefont {Nguyen}, \citenamefont {de-la
  Roza}, \citenamefont {Paulatto}, \citenamefont {Poncé}, \citenamefont
  {Rocca}, \citenamefont {Sabatini}, \citenamefont {Santra}, \citenamefont
  {Schlipf}, \citenamefont {Seitsonen}, \citenamefont {Smogunov}, \citenamefont
  {Timrov}, \citenamefont {Thonhauser}, \citenamefont {Umari}, \citenamefont
  {Vast}, \citenamefont {Wu},\ and\ \citenamefont {Baroni}}]{QE-2017}%
  \BibitemOpen
  \bibfield  {author} {\bibinfo {author} {\bibfnamefont {P.}~\bibnamefont
  {Giannozzi}}, \bibinfo {author} {\bibfnamefont {O.}~\bibnamefont
  {Andreussi}}, \bibinfo {author} {\bibfnamefont {T.}~\bibnamefont {Brumme}},
  \bibinfo {author} {\bibfnamefont {O.}~\bibnamefont {Bunau}}, \bibinfo
  {author} {\bibfnamefont {M.~B.}\ \bibnamefont {Nardelli}}, \bibinfo {author}
  {\bibfnamefont {M.}~\bibnamefont {Calandra}}, \bibinfo {author}
  {\bibfnamefont {R.}~\bibnamefont {Car}}, \bibinfo {author} {\bibfnamefont
  {C.}~\bibnamefont {Cavazzoni}}, \bibinfo {author} {\bibfnamefont
  {D.}~\bibnamefont {Ceresoli}}, \bibinfo {author} {\bibfnamefont
  {M.}~\bibnamefont {Cococcioni}}, \bibinfo {author} {\bibfnamefont
  {N.}~\bibnamefont {Colonna}}, \bibinfo {author} {\bibfnamefont
  {I.}~\bibnamefont {Carnimeo}}, \bibinfo {author} {\bibfnamefont {A.~D.}\
  \bibnamefont {Corso}}, \bibinfo {author} {\bibfnamefont {S.}~\bibnamefont
  {de~Gironcoli}}, \bibinfo {author} {\bibfnamefont {P.}~\bibnamefont
  {Delugas}}, \bibinfo {author} {\bibfnamefont {R.~A.~D.}\ \bibnamefont {Jr}},
  \bibinfo {author} {\bibfnamefont {A.}~\bibnamefont {Ferretti}}, \bibinfo
  {author} {\bibfnamefont {A.}~\bibnamefont {Floris}}, \bibinfo {author}
  {\bibfnamefont {G.}~\bibnamefont {Fratesi}}, \bibinfo {author} {\bibfnamefont
  {G.}~\bibnamefont {Fugallo}}, \bibinfo {author} {\bibfnamefont
  {R.}~\bibnamefont {Gebauer}}, \bibinfo {author} {\bibfnamefont
  {U.}~\bibnamefont {Gerstmann}}, \bibinfo {author} {\bibfnamefont
  {F.}~\bibnamefont {Giustino}}, \bibinfo {author} {\bibfnamefont
  {T.}~\bibnamefont {Gorni}}, \bibinfo {author} {\bibfnamefont
  {J.}~\bibnamefont {Jia}}, \bibinfo {author} {\bibfnamefont {M.}~\bibnamefont
  {Kawamura}}, \bibinfo {author} {\bibfnamefont {H.-Y.}\ \bibnamefont {Ko}},
  \bibinfo {author} {\bibfnamefont {A.}~\bibnamefont {Kokalj}}, \bibinfo
  {author} {\bibfnamefont {E.}~\bibnamefont {Küçükbenli}}, \bibinfo {author}
  {\bibfnamefont {M.}~\bibnamefont {Lazzeri}}, \bibinfo {author} {\bibfnamefont
  {M.}~\bibnamefont {Marsili}}, \bibinfo {author} {\bibfnamefont
  {N.}~\bibnamefont {Marzari}}, \bibinfo {author} {\bibfnamefont
  {F.}~\bibnamefont {Mauri}}, \bibinfo {author} {\bibfnamefont {N.~L.}\
  \bibnamefont {Nguyen}}, \bibinfo {author} {\bibfnamefont {H.-V.}\
  \bibnamefont {Nguyen}}, \bibinfo {author} {\bibfnamefont {A.~O.}\
  \bibnamefont {de-la Roza}}, \bibinfo {author} {\bibfnamefont
  {L.}~\bibnamefont {Paulatto}}, \bibinfo {author} {\bibfnamefont
  {S.}~\bibnamefont {Poncé}}, \bibinfo {author} {\bibfnamefont
  {D.}~\bibnamefont {Rocca}}, \bibinfo {author} {\bibfnamefont
  {R.}~\bibnamefont {Sabatini}}, \bibinfo {author} {\bibfnamefont
  {B.}~\bibnamefont {Santra}}, \bibinfo {author} {\bibfnamefont
  {M.}~\bibnamefont {Schlipf}}, \bibinfo {author} {\bibfnamefont {A.~P.}\
  \bibnamefont {Seitsonen}}, \bibinfo {author} {\bibfnamefont {A.}~\bibnamefont
  {Smogunov}}, \bibinfo {author} {\bibfnamefont {I.}~\bibnamefont {Timrov}},
  \bibinfo {author} {\bibfnamefont {T.}~\bibnamefont {Thonhauser}}, \bibinfo
  {author} {\bibfnamefont {P.}~\bibnamefont {Umari}}, \bibinfo {author}
  {\bibfnamefont {N.}~\bibnamefont {Vast}}, \bibinfo {author} {\bibfnamefont
  {X.}~\bibnamefont {Wu}}, \ and\ \bibinfo {author} {\bibfnamefont
  {S.}~\bibnamefont {Baroni}},\ }\href
  {http://stacks.iop.org/0953-8984/29/i=46/a=465901} {\bibfield  {journal}
  {\bibinfo  {journal} {Journal of Physics: Condensed Matter}\ }\textbf
  {\bibinfo {volume} {29}},\ \bibinfo {pages} {465901} (\bibinfo {year}
  {2017})}\BibitemShut {NoStop}%
\bibitem [{\citenamefont {Lejaeghere}\ \emph {et~al.}(2016)\citenamefont
  {Lejaeghere}, \citenamefont {Bihlmayer}, \citenamefont {Bj{\"o}rkman},
  \citenamefont {Blaha}, \citenamefont {Bl{\"u}gel}, \citenamefont {Blum},
  \citenamefont {Caliste}, \citenamefont {Castelli}, \citenamefont {Clark},
  \citenamefont {Dal~Corso}, \citenamefont {de~Gironcoli}, \citenamefont
  {Deutsch}, \citenamefont {Dewhurst}, \citenamefont {Di~Marco}, \citenamefont
  {Draxl}, \citenamefont {Du{\l}ak}, \citenamefont {Eriksson}, \citenamefont
  {Flores-Livas}, \citenamefont {Garrity}, \citenamefont {Genovese},
  \citenamefont {Giannozzi}, \citenamefont {Giantomassi}, \citenamefont
  {Goedecker}, \citenamefont {Gonze}, \citenamefont {Gr{\r a}n{\"a}s},
  \citenamefont {Gross}, \citenamefont {Gulans}, \citenamefont {Gygi},
  \citenamefont {Hamann}, \citenamefont {Hasnip}, \citenamefont {Holzwarth},
  \citenamefont {Iu{\c s}an}, \citenamefont {Jochym}, \citenamefont {Jollet},
  \citenamefont {Jones}, \citenamefont {Kresse}, \citenamefont {Koepernik},
  \citenamefont {K{\"u}{\c c}{\"u}kbenli}, \citenamefont {Kvashnin},
  \citenamefont {Locht}, \citenamefont {Lubeck}, \citenamefont {Marsman},
  \citenamefont {Marzari}, \citenamefont {Nitzsche}, \citenamefont
  {Nordstr{\"o}m}, \citenamefont {Ozaki}, \citenamefont {Paulatto},
  \citenamefont {Pickard}, \citenamefont {Poelmans}, \citenamefont {Probert},
  \citenamefont {Refson}, \citenamefont {Richter}, \citenamefont {Rignanese},
  \citenamefont {Saha}, \citenamefont {Scheffler}, \citenamefont {Schlipf},
  \citenamefont {Schwarz}, \citenamefont {Sharma}, \citenamefont {Tavazza},
  \citenamefont {Thunstr{\"o}m}, \citenamefont {Tkatchenko}, \citenamefont
  {Torrent}, \citenamefont {Vanderbilt}, \citenamefont {van Setten},
  \citenamefont {Van~Speybroeck}, \citenamefont {Wills}, \citenamefont {Yates},
  \citenamefont {Zhang},\ and\ \citenamefont {Cottenier}}]{Lejaeghere2016}%
  \BibitemOpen
  \bibfield  {author} {\bibinfo {author} {\bibfnamefont {K.}~\bibnamefont
  {Lejaeghere}}, \bibinfo {author} {\bibfnamefont {G.}~\bibnamefont
  {Bihlmayer}}, \bibinfo {author} {\bibfnamefont {T.}~\bibnamefont
  {Bj{\"o}rkman}}, \bibinfo {author} {\bibfnamefont {P.}~\bibnamefont {Blaha}},
  \bibinfo {author} {\bibfnamefont {S.}~\bibnamefont {Bl{\"u}gel}}, \bibinfo
  {author} {\bibfnamefont {V.}~\bibnamefont {Blum}}, \bibinfo {author}
  {\bibfnamefont {D.}~\bibnamefont {Caliste}}, \bibinfo {author} {\bibfnamefont
  {I.~E.}\ \bibnamefont {Castelli}}, \bibinfo {author} {\bibfnamefont {S.~J.}\
  \bibnamefont {Clark}}, \bibinfo {author} {\bibfnamefont {A.}~\bibnamefont
  {Dal~Corso}}, \bibinfo {author} {\bibfnamefont {S.}~\bibnamefont
  {de~Gironcoli}}, \bibinfo {author} {\bibfnamefont {T.}~\bibnamefont
  {Deutsch}}, \bibinfo {author} {\bibfnamefont {J.~K.}\ \bibnamefont
  {Dewhurst}}, \bibinfo {author} {\bibfnamefont {I.}~\bibnamefont {Di~Marco}},
  \bibinfo {author} {\bibfnamefont {C.}~\bibnamefont {Draxl}}, \bibinfo
  {author} {\bibfnamefont {M.}~\bibnamefont {Du{\l}ak}}, \bibinfo {author}
  {\bibfnamefont {O.}~\bibnamefont {Eriksson}}, \bibinfo {author}
  {\bibfnamefont {J.~A.}\ \bibnamefont {Flores-Livas}}, \bibinfo {author}
  {\bibfnamefont {K.~F.}\ \bibnamefont {Garrity}}, \bibinfo {author}
  {\bibfnamefont {L.}~\bibnamefont {Genovese}}, \bibinfo {author}
  {\bibfnamefont {P.}~\bibnamefont {Giannozzi}}, \bibinfo {author}
  {\bibfnamefont {M.}~\bibnamefont {Giantomassi}}, \bibinfo {author}
  {\bibfnamefont {S.}~\bibnamefont {Goedecker}}, \bibinfo {author}
  {\bibfnamefont {X.}~\bibnamefont {Gonze}}, \bibinfo {author} {\bibfnamefont
  {O.}~\bibnamefont {Gr{\r a}n{\"a}s}}, \bibinfo {author} {\bibfnamefont
  {E.~K.~U.}\ \bibnamefont {Gross}}, \bibinfo {author} {\bibfnamefont
  {A.}~\bibnamefont {Gulans}}, \bibinfo {author} {\bibfnamefont
  {F.}~\bibnamefont {Gygi}}, \bibinfo {author} {\bibfnamefont {D.~R.}\
  \bibnamefont {Hamann}}, \bibinfo {author} {\bibfnamefont {P.~J.}\
  \bibnamefont {Hasnip}}, \bibinfo {author} {\bibfnamefont {N.~A.~W.}\
  \bibnamefont {Holzwarth}}, \bibinfo {author} {\bibfnamefont {D.}~\bibnamefont
  {Iu{\c s}an}}, \bibinfo {author} {\bibfnamefont {D.~B.}\ \bibnamefont
  {Jochym}}, \bibinfo {author} {\bibfnamefont {F.}~\bibnamefont {Jollet}},
  \bibinfo {author} {\bibfnamefont {D.}~\bibnamefont {Jones}}, \bibinfo
  {author} {\bibfnamefont {G.}~\bibnamefont {Kresse}}, \bibinfo {author}
  {\bibfnamefont {K.}~\bibnamefont {Koepernik}}, \bibinfo {author}
  {\bibfnamefont {E.}~\bibnamefont {K{\"u}{\c c}{\"u}kbenli}}, \bibinfo
  {author} {\bibfnamefont {Y.~O.}\ \bibnamefont {Kvashnin}}, \bibinfo {author}
  {\bibfnamefont {I.~L.~M.}\ \bibnamefont {Locht}}, \bibinfo {author}
  {\bibfnamefont {S.}~\bibnamefont {Lubeck}}, \bibinfo {author} {\bibfnamefont
  {M.}~\bibnamefont {Marsman}}, \bibinfo {author} {\bibfnamefont
  {N.}~\bibnamefont {Marzari}}, \bibinfo {author} {\bibfnamefont
  {U.}~\bibnamefont {Nitzsche}}, \bibinfo {author} {\bibfnamefont
  {L.}~\bibnamefont {Nordstr{\"o}m}}, \bibinfo {author} {\bibfnamefont
  {T.}~\bibnamefont {Ozaki}}, \bibinfo {author} {\bibfnamefont
  {L.}~\bibnamefont {Paulatto}}, \bibinfo {author} {\bibfnamefont {C.~J.}\
  \bibnamefont {Pickard}}, \bibinfo {author} {\bibfnamefont {W.}~\bibnamefont
  {Poelmans}}, \bibinfo {author} {\bibfnamefont {M.~I.~J.}\ \bibnamefont
  {Probert}}, \bibinfo {author} {\bibfnamefont {K.}~\bibnamefont {Refson}},
  \bibinfo {author} {\bibfnamefont {M.}~\bibnamefont {Richter}}, \bibinfo
  {author} {\bibfnamefont {G.-M.}\ \bibnamefont {Rignanese}}, \bibinfo {author}
  {\bibfnamefont {S.}~\bibnamefont {Saha}}, \bibinfo {author} {\bibfnamefont
  {M.}~\bibnamefont {Scheffler}}, \bibinfo {author} {\bibfnamefont
  {M.}~\bibnamefont {Schlipf}}, \bibinfo {author} {\bibfnamefont
  {K.}~\bibnamefont {Schwarz}}, \bibinfo {author} {\bibfnamefont
  {S.}~\bibnamefont {Sharma}}, \bibinfo {author} {\bibfnamefont
  {F.}~\bibnamefont {Tavazza}}, \bibinfo {author} {\bibfnamefont
  {P.}~\bibnamefont {Thunstr{\"o}m}}, \bibinfo {author} {\bibfnamefont
  {A.}~\bibnamefont {Tkatchenko}}, \bibinfo {author} {\bibfnamefont
  {M.}~\bibnamefont {Torrent}}, \bibinfo {author} {\bibfnamefont
  {D.}~\bibnamefont {Vanderbilt}}, \bibinfo {author} {\bibfnamefont {M.~J.}\
  \bibnamefont {van Setten}}, \bibinfo {author} {\bibfnamefont
  {V.}~\bibnamefont {Van~Speybroeck}}, \bibinfo {author} {\bibfnamefont
  {J.~M.}\ \bibnamefont {Wills}}, \bibinfo {author} {\bibfnamefont {J.~R.}\
  \bibnamefont {Yates}}, \bibinfo {author} {\bibfnamefont {G.-X.}\ \bibnamefont
  {Zhang}}, \ and\ \bibinfo {author} {\bibfnamefont {S.}~\bibnamefont
  {Cottenier}},\ }\href {\doibase 10.1126/science.aad3000} {\bibfield
  {journal} {\bibinfo  {journal} {Science}\ }\textbf {\bibinfo {volume} {351}}
  (\bibinfo {year} {2016}),\ 10.1126/science.aad3000},\ \Eprint
  {http://arxiv.org/abs/https://science.sciencemag.org/content/351/6280/aad3000.full.pdf}
  {https://science.sciencemag.org/content/351/6280/aad3000.full.pdf}
  \BibitemShut {NoStop}%
\bibitem [{\citenamefont {Prandini}\ \emph {et~al.}(2018)\citenamefont
  {Prandini}, \citenamefont {Marrazzo}, \citenamefont {Castelli}, \citenamefont
  {Mounet},\ and\ \citenamefont {Marzari}}]{Prandini2018}%
  \BibitemOpen
  \bibfield  {author} {\bibinfo {author} {\bibfnamefont {G.}~\bibnamefont
  {Prandini}}, \bibinfo {author} {\bibfnamefont {A.}~\bibnamefont {Marrazzo}},
  \bibinfo {author} {\bibfnamefont {I.~E.}\ \bibnamefont {Castelli}}, \bibinfo
  {author} {\bibfnamefont {N.}~\bibnamefont {Mounet}}, \ and\ \bibinfo {author}
  {\bibfnamefont {N.}~\bibnamefont {Marzari}},\ }\href {\doibase
  10.1038/s41524-018-0127-2} {\bibfield  {journal} {\bibinfo  {journal} {npj
  Computational Materials}\ }\textbf {\bibinfo {volume} {4}},\ \bibinfo {pages}
  {72} (\bibinfo {year} {2018})}\BibitemShut {NoStop}%
\bibitem [{\citenamefont {Taroni}\ \emph {et~al.}(2008)\citenamefont {Taroni},
  \citenamefont {Bramwell},\ and\ \citenamefont {Holdsworth}}]{Taroni2008}%
  \BibitemOpen
  \bibfield  {author} {\bibinfo {author} {\bibfnamefont {A.}~\bibnamefont
  {Taroni}}, \bibinfo {author} {\bibfnamefont {S.~T.}\ \bibnamefont
  {Bramwell}}, \ and\ \bibinfo {author} {\bibfnamefont {P.~C.~W.}\ \bibnamefont
  {Holdsworth}},\ }\href {\doibase 10.1088/0953-8984/20/27/275233} {\bibfield
  {journal} {\bibinfo  {journal} {Journal of Physics: Condensed Matter}\
  }\textbf {\bibinfo {volume} {20}},\ \bibinfo {pages} {275233} (\bibinfo
  {year} {2008})}\BibitemShut {NoStop}%
\bibitem [{\citenamefont {Lane}\ \emph {et~al.}(2021)\citenamefont {Lane},
  \citenamefont {Pachoud}, \citenamefont {Rodriguez-Rivera}, \citenamefont
  {Songvilay}, \citenamefont {Xu}, \citenamefont {Gehring}, \citenamefont
  {Attfield}, \citenamefont {Ewings},\ and\ \citenamefont {Stock}}]{Lane2021}%
  \BibitemOpen
  \bibfield  {author} {\bibinfo {author} {\bibfnamefont {H.}~\bibnamefont
  {Lane}}, \bibinfo {author} {\bibfnamefont {E.}~\bibnamefont {Pachoud}},
  \bibinfo {author} {\bibfnamefont {J.~A.}\ \bibnamefont {Rodriguez-Rivera}},
  \bibinfo {author} {\bibfnamefont {M.}~\bibnamefont {Songvilay}}, \bibinfo
  {author} {\bibfnamefont {G.}~\bibnamefont {Xu}}, \bibinfo {author}
  {\bibfnamefont {P.~M.}\ \bibnamefont {Gehring}}, \bibinfo {author}
  {\bibfnamefont {J.~P.}\ \bibnamefont {Attfield}}, \bibinfo {author}
  {\bibfnamefont {R.~A.}\ \bibnamefont {Ewings}}, \ and\ \bibinfo {author}
  {\bibfnamefont {C.}~\bibnamefont {Stock}},\ }\href {\doibase
  10.1103/PhysRevB.104.L020411} {\bibfield  {journal} {\bibinfo  {journal}
  {Phys. Rev. B}\ }\textbf {\bibinfo {volume} {104}},\ \bibinfo {pages}
  {L020411} (\bibinfo {year} {2021})}\BibitemShut {NoStop}%
\bibitem [{\citenamefont {Baker}\ \emph {et~al.}(2005)\citenamefont {Baker},
  \citenamefont {Lancaster}, \citenamefont {Blundell}, \citenamefont {Brooks},
  \citenamefont {Hayes}, \citenamefont {Prabhakaran},\ and\ \citenamefont
  {Pratt}}]{Baker2005}%
  \BibitemOpen
  \bibfield  {author} {\bibinfo {author} {\bibfnamefont {P.~J.}\ \bibnamefont
  {Baker}}, \bibinfo {author} {\bibfnamefont {T.}~\bibnamefont {Lancaster}},
  \bibinfo {author} {\bibfnamefont {S.~J.}\ \bibnamefont {Blundell}}, \bibinfo
  {author} {\bibfnamefont {M.~L.}\ \bibnamefont {Brooks}}, \bibinfo {author}
  {\bibfnamefont {W.}~\bibnamefont {Hayes}}, \bibinfo {author} {\bibfnamefont
  {D.}~\bibnamefont {Prabhakaran}}, \ and\ \bibinfo {author} {\bibfnamefont
  {F.~L.}\ \bibnamefont {Pratt}},\ }\href {\doibase 10.1103/PhysRevB.72.104414}
  {\bibfield  {journal} {\bibinfo  {journal} {Phys. Rev. B}\ }\textbf {\bibinfo
  {volume} {72}},\ \bibinfo {pages} {104414} (\bibinfo {year}
  {2005})}\BibitemShut {NoStop}%
\bibitem [{\citenamefont {Forslund}\ \emph
  {et~al.}(2020{\natexlab{a}})\citenamefont {Forslund}, \citenamefont {Ohta},
  \citenamefont {Kamazawa}, \citenamefont {Stubbs}, \citenamefont {Ofer},
  \citenamefont {M\aa{}nsson}, \citenamefont {Michioka}, \citenamefont
  {Yoshimura}, \citenamefont {Hitti}, \citenamefont {Arseneau}, \citenamefont
  {Morris}, \citenamefont {Ansaldo}, \citenamefont {Brewer},\ and\
  \citenamefont {Sugiyama}}]{Forslund_2020_Na}%
  \BibitemOpen
  \bibfield  {author} {\bibinfo {author} {\bibfnamefont {O.~K.}\ \bibnamefont
  {Forslund}}, \bibinfo {author} {\bibfnamefont {H.}~\bibnamefont {Ohta}},
  \bibinfo {author} {\bibfnamefont {K.}~\bibnamefont {Kamazawa}}, \bibinfo
  {author} {\bibfnamefont {S.~L.}\ \bibnamefont {Stubbs}}, \bibinfo {author}
  {\bibfnamefont {O.}~\bibnamefont {Ofer}}, \bibinfo {author} {\bibfnamefont
  {M.}~\bibnamefont {M\aa{}nsson}}, \bibinfo {author} {\bibfnamefont
  {C.}~\bibnamefont {Michioka}}, \bibinfo {author} {\bibfnamefont
  {K.}~\bibnamefont {Yoshimura}}, \bibinfo {author} {\bibfnamefont
  {B.}~\bibnamefont {Hitti}}, \bibinfo {author} {\bibfnamefont
  {D.}~\bibnamefont {Arseneau}}, \bibinfo {author} {\bibfnamefont {G.~D.}\
  \bibnamefont {Morris}}, \bibinfo {author} {\bibfnamefont {E.~J.}\
  \bibnamefont {Ansaldo}}, \bibinfo {author} {\bibfnamefont {J.~H.}\
  \bibnamefont {Brewer}}, \ and\ \bibinfo {author} {\bibfnamefont
  {J.}~\bibnamefont {Sugiyama}},\ }\href {\doibase 10.1103/PhysRevB.102.184412}
  {\bibfield  {journal} {\bibinfo  {journal} {Phys. Rev. B}\ }\textbf {\bibinfo
  {volume} {102}},\ \bibinfo {pages} {184412} (\bibinfo {year}
  {2020}{\natexlab{a}})}\BibitemShut {NoStop}%
\bibitem [{\citenamefont {Forslund}\ \emph
  {et~al.}(2021{\natexlab{b}})\citenamefont {Forslund}, \citenamefont
  {Andreica}, \citenamefont {Ohta}, \citenamefont {Imai}, \citenamefont
  {Michioka}, \citenamefont {Yoshimura}, \citenamefont {M{\aa}nsson},\ and\
  \citenamefont {Sugiyama}}]{Forslund_2021_La}%
  \BibitemOpen
  \bibfield  {author} {\bibinfo {author} {\bibfnamefont {O.~K.}\ \bibnamefont
  {Forslund}}, \bibinfo {author} {\bibfnamefont {D.}~\bibnamefont {Andreica}},
  \bibinfo {author} {\bibfnamefont {H.}~\bibnamefont {Ohta}}, \bibinfo {author}
  {\bibfnamefont {M.}~\bibnamefont {Imai}}, \bibinfo {author} {\bibfnamefont
  {C.}~\bibnamefont {Michioka}}, \bibinfo {author} {\bibfnamefont
  {K.}~\bibnamefont {Yoshimura}}, \bibinfo {author} {\bibfnamefont
  {M.}~\bibnamefont {M{\aa}nsson}}, \ and\ \bibinfo {author} {\bibfnamefont
  {J.}~\bibnamefont {Sugiyama}},\ }\href {\doibase 10.1088/1402-4896/ac3cf9}
  {\bibfield  {journal} {\bibinfo  {journal} {Physica Scripta}\ }\textbf
  {\bibinfo {volume} {96}},\ \bibinfo {pages} {125864} (\bibinfo {year}
  {2021}{\natexlab{b}})}\BibitemShut {NoStop}%
\bibitem [{\citenamefont {Bonfà}\ \emph {et~al.}()\citenamefont {Bonfà},
  \citenamefont {Onuorah},\ and\ \citenamefont {Renzi}}]{Bonfa2017}%
  \BibitemOpen
  \bibfield  {author} {\bibinfo {author} {\bibfnamefont {P.}~\bibnamefont
  {Bonfà}}, \bibinfo {author} {\bibfnamefont {I.~J.}\ \bibnamefont {Onuorah}},
  \ and\ \bibinfo {author} {\bibfnamefont {R.~D.}\ \bibnamefont {Renzi}},\
  }\enquote {\bibinfo {title} {Introduction and a quick look at muesr, the
  magnetic structure and muon embedding site refinement suite},}\ in\ \href
  {\doibase 10.7566/JPSCP.21.011052} {\emph {\bibinfo {booktitle} {Proceedings
  of the 14th International Conference on Muon Spin Rotation, Relaxation and
  Resonance (SR2017)}}},\ \Eprint
  {http://arxiv.org/abs/https://journals.jps.jp/doi/pdf/10.7566/JPSCP.21.011052}
  {https://journals.jps.jp/doi/pdf/10.7566/JPSCP.21.011052} \BibitemShut
  {NoStop}%
\bibitem [{\citenamefont {Nozaki}\ \emph {et~al.}(2018)\citenamefont {Nozaki},
  \citenamefont {Sakurai}, \citenamefont {Ofer}, \citenamefont {Ansaldo},
  \citenamefont {Brewer}, \citenamefont {Chow}, \citenamefont {Pomjakushin},
  \citenamefont {Keller}, \citenamefont {Prša}, \citenamefont {Miwa},
  \citenamefont {Månsson},\ and\ \citenamefont {Sugiyama}}]{Nozaki2018}%
  \BibitemOpen
  \bibfield  {author} {\bibinfo {author} {\bibfnamefont {H.}~\bibnamefont
  {Nozaki}}, \bibinfo {author} {\bibfnamefont {H.}~\bibnamefont {Sakurai}},
  \bibinfo {author} {\bibfnamefont {O.}~\bibnamefont {Ofer}}, \bibinfo {author}
  {\bibfnamefont {E.~J.}\ \bibnamefont {Ansaldo}}, \bibinfo {author}
  {\bibfnamefont {J.~H.}\ \bibnamefont {Brewer}}, \bibinfo {author}
  {\bibfnamefont {K.~H.}\ \bibnamefont {Chow}}, \bibinfo {author}
  {\bibfnamefont {V.}~\bibnamefont {Pomjakushin}}, \bibinfo {author}
  {\bibfnamefont {L.}~\bibnamefont {Keller}}, \bibinfo {author} {\bibfnamefont
  {K.}~\bibnamefont {Prša}}, \bibinfo {author} {\bibfnamefont
  {K.}~\bibnamefont {Miwa}}, \bibinfo {author} {\bibfnamefont {M.}~\bibnamefont
  {Månsson}}, \ and\ \bibinfo {author} {\bibfnamefont {J.}~\bibnamefont
  {Sugiyama}},\ }\href {\doibase https://doi.org/10.1016/j.physb.2017.11.011}
  {\bibfield  {journal} {\bibinfo  {journal} {Physica B: Condensed Matter}\
  }\textbf {\bibinfo {volume} {551}},\ \bibinfo {pages} {137} (\bibinfo {year}
  {2018})},\ \bibinfo {note} {the 11th International Conference on Neutron
  Scattering (ICNS 2017)}\BibitemShut {NoStop}%
\bibitem [{\citenamefont {Potashnikov}\ \emph {et~al.}(2021)\citenamefont
  {Potashnikov}, \citenamefont {Caspi}, \citenamefont {Pesach}, \citenamefont
  {Tao}, \citenamefont {Rosen}, \citenamefont {Sheptyakov}, \citenamefont
  {Evans}, \citenamefont {Ritter}, \citenamefont {Salman}, \citenamefont
  {Bonfa}, \citenamefont {Ouisse}, \citenamefont {Barbier}, \citenamefont
  {Rivin},\ and\ \citenamefont {Keren}}]{Potashnikov2021}%
  \BibitemOpen
  \bibfield  {author} {\bibinfo {author} {\bibfnamefont {D.}~\bibnamefont
  {Potashnikov}}, \bibinfo {author} {\bibfnamefont {E.~N.}\ \bibnamefont
  {Caspi}}, \bibinfo {author} {\bibfnamefont {A.}~\bibnamefont {Pesach}},
  \bibinfo {author} {\bibfnamefont {Q.}~\bibnamefont {Tao}}, \bibinfo {author}
  {\bibfnamefont {J.}~\bibnamefont {Rosen}}, \bibinfo {author} {\bibfnamefont
  {D.}~\bibnamefont {Sheptyakov}}, \bibinfo {author} {\bibfnamefont {H.~A.}\
  \bibnamefont {Evans}}, \bibinfo {author} {\bibfnamefont {C.}~\bibnamefont
  {Ritter}}, \bibinfo {author} {\bibfnamefont {Z.}~\bibnamefont {Salman}},
  \bibinfo {author} {\bibfnamefont {P.}~\bibnamefont {Bonfa}}, \bibinfo
  {author} {\bibfnamefont {T.}~\bibnamefont {Ouisse}}, \bibinfo {author}
  {\bibfnamefont {M.}~\bibnamefont {Barbier}}, \bibinfo {author} {\bibfnamefont
  {O.}~\bibnamefont {Rivin}}, \ and\ \bibinfo {author} {\bibfnamefont
  {A.}~\bibnamefont {Keren}},\ }\href {\doibase 10.1103/PhysRevB.104.174440}
  {\bibfield  {journal} {\bibinfo  {journal} {Phys. Rev. B}\ }\textbf {\bibinfo
  {volume} {104}},\ \bibinfo {pages} {174440} (\bibinfo {year}
  {2021})}\BibitemShut {NoStop}%
\bibitem [{\citenamefont {Sulaiman}\ \emph {et~al.}(1994)\citenamefont
  {Sulaiman}, \citenamefont {Srinivas}, \citenamefont {Sahoo}, \citenamefont
  {Hagelberg}, \citenamefont {Das}, \citenamefont {Torikai},\ and\
  \citenamefont {Nagamine}}]{Sulaiman1994}%
  \BibitemOpen
  \bibfield  {author} {\bibinfo {author} {\bibfnamefont {S.~B.}\ \bibnamefont
  {Sulaiman}}, \bibinfo {author} {\bibfnamefont {S.}~\bibnamefont {Srinivas}},
  \bibinfo {author} {\bibfnamefont {N.}~\bibnamefont {Sahoo}}, \bibinfo
  {author} {\bibfnamefont {F.}~\bibnamefont {Hagelberg}}, \bibinfo {author}
  {\bibfnamefont {T.~P.}\ \bibnamefont {Das}}, \bibinfo {author} {\bibfnamefont
  {E.}~\bibnamefont {Torikai}}, \ and\ \bibinfo {author} {\bibfnamefont
  {K.}~\bibnamefont {Nagamine}},\ }\href {\doibase 10.1103/PhysRevB.49.9879}
  {\bibfield  {journal} {\bibinfo  {journal} {Phys. Rev. B}\ }\textbf {\bibinfo
  {volume} {49}},\ \bibinfo {pages} {9879} (\bibinfo {year}
  {1994})}\BibitemShut {NoStop}%
\bibitem [{\citenamefont {Forslund}\ \emph
  {et~al.}(2020{\natexlab{b}})\citenamefont {Forslund}, \citenamefont
  {Papadopoulos}, \citenamefont {Nocerino}, \citenamefont {Morris},
  \citenamefont {Hitti}, \citenamefont {Arseneau}, \citenamefont {Pomjakushin},
  \citenamefont {Matsubara}, \citenamefont {Orain}, \citenamefont {Svedlindh},
  \citenamefont {Andreica}, \citenamefont {Jana}, \citenamefont {Sugiyama},
  \citenamefont {M\aa{}nsson},\ and\ \citenamefont {Sassa}}]{Forslund2020_La}%
  \BibitemOpen
  \bibfield  {author} {\bibinfo {author} {\bibfnamefont {O.~K.}\ \bibnamefont
  {Forslund}}, \bibinfo {author} {\bibfnamefont {K.}~\bibnamefont
  {Papadopoulos}}, \bibinfo {author} {\bibfnamefont {E.}~\bibnamefont
  {Nocerino}}, \bibinfo {author} {\bibfnamefont {G.}~\bibnamefont {Morris}},
  \bibinfo {author} {\bibfnamefont {B.}~\bibnamefont {Hitti}}, \bibinfo
  {author} {\bibfnamefont {D.}~\bibnamefont {Arseneau}}, \bibinfo {author}
  {\bibfnamefont {V.}~\bibnamefont {Pomjakushin}}, \bibinfo {author}
  {\bibfnamefont {N.}~\bibnamefont {Matsubara}}, \bibinfo {author}
  {\bibfnamefont {J.-C.}\ \bibnamefont {Orain}}, \bibinfo {author}
  {\bibfnamefont {P.}~\bibnamefont {Svedlindh}}, \bibinfo {author}
  {\bibfnamefont {D.}~\bibnamefont {Andreica}}, \bibinfo {author}
  {\bibfnamefont {S.}~\bibnamefont {Jana}}, \bibinfo {author} {\bibfnamefont
  {J.}~\bibnamefont {Sugiyama}}, \bibinfo {author} {\bibfnamefont
  {M.}~\bibnamefont {M\aa{}nsson}}, \ and\ \bibinfo {author} {\bibfnamefont
  {Y.}~\bibnamefont {Sassa}},\ }\href {\doibase 10.1103/PhysRevB.102.144409}
  {\bibfield  {journal} {\bibinfo  {journal} {Phys. Rev. B}\ }\textbf {\bibinfo
  {volume} {102}},\ \bibinfo {pages} {144409} (\bibinfo {year}
  {2020}{\natexlab{b}})}\BibitemShut {NoStop}%
\bibitem [{\citenamefont {Yang}\ \emph {et~al.}(2020)\citenamefont {Yang},
  \citenamefont {Fan}, \citenamefont {Wang}, \citenamefont {Khomskii},\ and\
  \citenamefont {Wu}}]{Yang2020}%
  \BibitemOpen
  \bibfield  {author} {\bibinfo {author} {\bibfnamefont {K.}~\bibnamefont
  {Yang}}, \bibinfo {author} {\bibfnamefont {F.}~\bibnamefont {Fan}}, \bibinfo
  {author} {\bibfnamefont {H.}~\bibnamefont {Wang}}, \bibinfo {author}
  {\bibfnamefont {D.~I.}\ \bibnamefont {Khomskii}}, \ and\ \bibinfo {author}
  {\bibfnamefont {H.}~\bibnamefont {Wu}},\ }\href {\doibase
  10.1103/PhysRevB.101.100402} {\bibfield  {journal} {\bibinfo  {journal}
  {Phys. Rev. B}\ }\textbf {\bibinfo {volume} {101}},\ \bibinfo {pages}
  {100402} (\bibinfo {year} {2020})}\BibitemShut {NoStop}%
\bibitem [{\citenamefont {Nguyen}\ \emph {et~al.}(2021)\citenamefont {Nguyen},
  \citenamefont {Yamauchi}, \citenamefont {Oguchi}, \citenamefont {Amoroso},\
  and\ \citenamefont {Picozzi}}]{Nguyen2021}%
  \BibitemOpen
  \bibfield  {author} {\bibinfo {author} {\bibfnamefont {T.~P.~T.}\
  \bibnamefont {Nguyen}}, \bibinfo {author} {\bibfnamefont {K.}~\bibnamefont
  {Yamauchi}}, \bibinfo {author} {\bibfnamefont {T.}~\bibnamefont {Oguchi}},
  \bibinfo {author} {\bibfnamefont {D.}~\bibnamefont {Amoroso}}, \ and\
  \bibinfo {author} {\bibfnamefont {S.}~\bibnamefont {Picozzi}},\ }\href
  {\doibase 10.1103/PhysRevB.104.014414} {\bibfield  {journal} {\bibinfo
  {journal} {Phys. Rev. B}\ }\textbf {\bibinfo {volume} {104}},\ \bibinfo
  {pages} {014414} (\bibinfo {year} {2021})}\BibitemShut {NoStop}%
\bibitem [{\citenamefont {Månsson}\ \emph {et~al.}(2012)\citenamefont
  {Månsson}, \citenamefont {Prša}, \citenamefont {Sugiyama}, \citenamefont
  {Nozaki}, \citenamefont {Amato}, \citenamefont {Omura}, \citenamefont
  {Kimura},\ and\ \citenamefont {Hagiwara}}]{Mansson2012}%
  \BibitemOpen
  \bibfield  {author} {\bibinfo {author} {\bibfnamefont {M.}~\bibnamefont
  {Månsson}}, \bibinfo {author} {\bibfnamefont {K.}~\bibnamefont {Prša}},
  \bibinfo {author} {\bibfnamefont {J.}~\bibnamefont {Sugiyama}}, \bibinfo
  {author} {\bibfnamefont {H.}~\bibnamefont {Nozaki}}, \bibinfo {author}
  {\bibfnamefont {A.}~\bibnamefont {Amato}}, \bibinfo {author} {\bibfnamefont
  {K.}~\bibnamefont {Omura}}, \bibinfo {author} {\bibfnamefont
  {S.}~\bibnamefont {Kimura}}, \ and\ \bibinfo {author} {\bibfnamefont
  {M.}~\bibnamefont {Hagiwara}},\ }\href {\doibase
  https://doi.org/10.1016/j.phpro.2012.04.060} {\bibfield  {journal} {\bibinfo
  {journal} {Physics Procedia}\ }\textbf {\bibinfo {volume} {30}},\ \bibinfo
  {pages} {146} (\bibinfo {year} {2012})},\ \bibinfo {note} {12th International
  Conference on Muon Spin Rotation, Relaxation and Resonance
  ($\mu$SR2011)}\BibitemShut {NoStop}%
\bibitem [{\citenamefont {Sugiyama}\ \emph {et~al.}(2005)\citenamefont
  {Sugiyama}, \citenamefont {Nozaki}, \citenamefont {Brewer}, \citenamefont
  {Ansaldo}, \citenamefont {Takami}, \citenamefont {Ikuta},\ and\ \citenamefont
  {Mizutani}}]{Sugiyama2005}%
  \BibitemOpen
  \bibfield  {author} {\bibinfo {author} {\bibfnamefont {J.}~\bibnamefont
  {Sugiyama}}, \bibinfo {author} {\bibfnamefont {H.}~\bibnamefont {Nozaki}},
  \bibinfo {author} {\bibfnamefont {J.~H.}\ \bibnamefont {Brewer}}, \bibinfo
  {author} {\bibfnamefont {E.~J.}\ \bibnamefont {Ansaldo}}, \bibinfo {author}
  {\bibfnamefont {T.}~\bibnamefont {Takami}}, \bibinfo {author} {\bibfnamefont
  {H.}~\bibnamefont {Ikuta}}, \ and\ \bibinfo {author} {\bibfnamefont
  {U.}~\bibnamefont {Mizutani}},\ }\href {\doibase 10.1103/PhysRevB.72.064418}
  {\bibfield  {journal} {\bibinfo  {journal} {Phys. Rev. B}\ }\textbf {\bibinfo
  {volume} {72}},\ \bibinfo {pages} {064418} (\bibinfo {year}
  {2005})}\BibitemShut {NoStop}%
\bibitem [{\citenamefont {Sugiyama}\ \emph {et~al.}(2006)\citenamefont
  {Sugiyama}, \citenamefont {Nozaki}, \citenamefont {Ikedo}, \citenamefont
  {Mukai}, \citenamefont {Andreica}, \citenamefont {Amato}, \citenamefont
  {Brewer}, \citenamefont {Ansaldo}, \citenamefont {Morris}, \citenamefont
  {Takami},\ and\ \citenamefont {Ikuta}}]{Sugiyama2006}%
  \BibitemOpen
  \bibfield  {author} {\bibinfo {author} {\bibfnamefont {J.}~\bibnamefont
  {Sugiyama}}, \bibinfo {author} {\bibfnamefont {H.}~\bibnamefont {Nozaki}},
  \bibinfo {author} {\bibfnamefont {Y.}~\bibnamefont {Ikedo}}, \bibinfo
  {author} {\bibfnamefont {K.}~\bibnamefont {Mukai}}, \bibinfo {author}
  {\bibfnamefont {D.}~\bibnamefont {Andreica}}, \bibinfo {author}
  {\bibfnamefont {A.}~\bibnamefont {Amato}}, \bibinfo {author} {\bibfnamefont
  {J.~H.}\ \bibnamefont {Brewer}}, \bibinfo {author} {\bibfnamefont {E.~J.}\
  \bibnamefont {Ansaldo}}, \bibinfo {author} {\bibfnamefont {G.~D.}\
  \bibnamefont {Morris}}, \bibinfo {author} {\bibfnamefont {T.}~\bibnamefont
  {Takami}}, \ and\ \bibinfo {author} {\bibfnamefont {H.}~\bibnamefont
  {Ikuta}},\ }\href {\doibase 10.1103/PhysRevLett.96.197206} {\bibfield
  {journal} {\bibinfo  {journal} {Phys. Rev. Lett.}\ }\textbf {\bibinfo
  {volume} {96}},\ \bibinfo {pages} {197206} (\bibinfo {year}
  {2006})}\BibitemShut {NoStop}%
\bibitem [{\citenamefont {Sugiyama}\ \emph {et~al.}(2008)\citenamefont
  {Sugiyama}, \citenamefont {Nozaki}, \citenamefont {Ikedo}, \citenamefont
  {Russo}, \citenamefont {Mukai}, \citenamefont {Andreica}, \citenamefont
  {Amato}, \citenamefont {Takami},\ and\ \citenamefont {Ikuta}}]{Sugiyama2008}%
  \BibitemOpen
  \bibfield  {author} {\bibinfo {author} {\bibfnamefont {J.}~\bibnamefont
  {Sugiyama}}, \bibinfo {author} {\bibfnamefont {H.}~\bibnamefont {Nozaki}},
  \bibinfo {author} {\bibfnamefont {Y.}~\bibnamefont {Ikedo}}, \bibinfo
  {author} {\bibfnamefont {P.~L.}\ \bibnamefont {Russo}}, \bibinfo {author}
  {\bibfnamefont {K.}~\bibnamefont {Mukai}}, \bibinfo {author} {\bibfnamefont
  {D.}~\bibnamefont {Andreica}}, \bibinfo {author} {\bibfnamefont
  {A.}~\bibnamefont {Amato}}, \bibinfo {author} {\bibfnamefont
  {T.}~\bibnamefont {Takami}}, \ and\ \bibinfo {author} {\bibfnamefont
  {H.}~\bibnamefont {Ikuta}},\ }\href {\doibase 10.1103/PhysRevB.77.092409}
  {\bibfield  {journal} {\bibinfo  {journal} {Phys. Rev. B}\ }\textbf {\bibinfo
  {volume} {77}},\ \bibinfo {pages} {092409} (\bibinfo {year}
  {2008})}\BibitemShut {NoStop}%
\bibitem [{\citenamefont {Schollw{\"o}ck}\ \emph {et~al.}(2008)\citenamefont
  {Schollw{\"o}ck}, \citenamefont {Richter}, \citenamefont {Farnell},\ and\
  \citenamefont {Bishop}}]{Schollwock2008}%
  \BibitemOpen
  \bibfield  {author} {\bibinfo {author} {\bibfnamefont {U.}~\bibnamefont
  {Schollw{\"o}ck}}, \bibinfo {author} {\bibfnamefont {J.}~\bibnamefont
  {Richter}}, \bibinfo {author} {\bibfnamefont {D.~J.}\ \bibnamefont
  {Farnell}}, \ and\ \bibinfo {author} {\bibfnamefont {R.~F.}\ \bibnamefont
  {Bishop}},\ }\href@noop {} {\emph {\bibinfo {title} {Quantum magnetism}}},\
  Vol.\ \bibinfo {volume} {645}\ (\bibinfo  {publisher} {Springer},\ \bibinfo
  {year} {2008})\BibitemShut {NoStop}%
\bibitem [{\citenamefont {He}\ \emph {et~al.}(2005)\citenamefont {He},
  \citenamefont {Fu}, \citenamefont {Ky{\^o}men}, \citenamefont {Taniyama},\
  and\ \citenamefont {Itoh}}]{He2005}%
  \BibitemOpen
  \bibfield  {author} {\bibinfo {author} {\bibfnamefont {Z.}~\bibnamefont
  {He}}, \bibinfo {author} {\bibfnamefont {D.}~\bibnamefont {Fu}}, \bibinfo
  {author} {\bibfnamefont {T.}~\bibnamefont {Ky{\^o}men}}, \bibinfo {author}
  {\bibfnamefont {T.}~\bibnamefont {Taniyama}}, \ and\ \bibinfo {author}
  {\bibfnamefont {M.}~\bibnamefont {Itoh}},\ }\href {\doibase
  10.1021/cm050760e} {\bibfield  {journal} {\bibinfo  {journal} {Chemistry of
  Materials}\ }\textbf {\bibinfo {volume} {17}},\ \bibinfo {pages} {2924}
  (\bibinfo {year} {2005})}\BibitemShut {NoStop}%
\bibitem [{\citenamefont {Yamauchi}\ \emph {et~al.}(2011)\citenamefont
  {Yamauchi}, \citenamefont {Ueda}, \citenamefont {Isobe},\ and\ \citenamefont
  {Ueda}}]{Yamauchi2011}%
  \BibitemOpen
  \bibfield  {author} {\bibinfo {author} {\bibfnamefont {T.}~\bibnamefont
  {Yamauchi}}, \bibinfo {author} {\bibfnamefont {H.}~\bibnamefont {Ueda}},
  \bibinfo {author} {\bibfnamefont {M.}~\bibnamefont {Isobe}}, \ and\ \bibinfo
  {author} {\bibfnamefont {Y.}~\bibnamefont {Ueda}},\ }\href {\doibase
  10.1103/PhysRevB.84.115104} {\bibfield  {journal} {\bibinfo  {journal} {Phys.
  Rev. B}\ }\textbf {\bibinfo {volume} {84}},\ \bibinfo {pages} {115104}
  (\bibinfo {year} {2011})}\BibitemShut {NoStop}%
\bibitem [{\citenamefont {McGuire}\ \emph {et~al.}(2017)\citenamefont
  {McGuire}, \citenamefont {Clark}, \citenamefont {KC}, \citenamefont {Chance},
  \citenamefont {Jellison}, \citenamefont {Cooper}, \citenamefont {Xu},\ and\
  \citenamefont {Sales}}]{McGuire2017}%
  \BibitemOpen
  \bibfield  {author} {\bibinfo {author} {\bibfnamefont {M.~A.}\ \bibnamefont
  {McGuire}}, \bibinfo {author} {\bibfnamefont {G.}~\bibnamefont {Clark}},
  \bibinfo {author} {\bibfnamefont {S.}~\bibnamefont {KC}}, \bibinfo {author}
  {\bibfnamefont {W.~M.}\ \bibnamefont {Chance}}, \bibinfo {author}
  {\bibfnamefont {G.~E.}\ \bibnamefont {Jellison}}, \bibinfo {author}
  {\bibfnamefont {V.~R.}\ \bibnamefont {Cooper}}, \bibinfo {author}
  {\bibfnamefont {X.}~\bibnamefont {Xu}}, \ and\ \bibinfo {author}
  {\bibfnamefont {B.~C.}\ \bibnamefont {Sales}},\ }\href {\doibase
  10.1103/PhysRevMaterials.1.014001} {\bibfield  {journal} {\bibinfo  {journal}
  {Phys. Rev. Materials}\ }\textbf {\bibinfo {volume} {1}},\ \bibinfo {pages}
  {014001} (\bibinfo {year} {2017})}\BibitemShut {NoStop}%
\end{thebibliography}%
\end{document}